\documentclass{article}
\usepackage[english]{babel}
\usepackage{microtype}

\usepackage[utf8]{inputenc}
\usepackage{amsmath}
\usepackage{graphicx}
\usepackage{multirow}
\usepackage{booktabs}
\usepackage{makecell}
\usepackage{ulem}
\usepackage{subcaption}
\usepackage{tikz}
\usepackage{feynmp-auto}
\usepackage[compat=1.1.0]{tikz-feynman}
\usepackage{float}
\usepackage{xcolor} 
\usepackage{amssymb}
\usepackage{dsfont}

\def\bea{\begin{eqnarray} }
\def\eea{ \end{eqnarray}}

\newcommand{\eq}[1]{Eq.~\eqref{#1}}
\newcommand{\fig}[1]{Fig.~\ref{#1}}
\newcommand{\tab}[1]{Tab.~\ref{#1}}

\newcommand{\Mp}{M_{\text{Pl}}}

\newcommand{\rsw}{\rm sw.}
\newcommand{\turb}{\rm turb.}
\usepackage{ulem}

\usepackage{jheppub}

\DeclareMathOperator*{\Sumint}{%
\mathchoice%
  {\ooalign{$\displaystyle\sum$\cr\hidewidth$\displaystyle\int$\hidewidth\cr}}
  {\ooalign{\raisebox{.14\height}{\scalebox{.7}{$\textstyle\sum$}}\cr\hidewidth$\textstyle\int$\hidewidth\cr}}
  {\ooalign{\raisebox{.2\height}{\scalebox{.6}{$\scriptstyle\sum$}}\cr$\scriptstyle\int$\cr}}
  {\ooalign{\raisebox{.2\height}{\scalebox{.6}{$\scriptstyle\sum$}}\cr$\scriptstyle\int$\cr}}
}

\usepackage{etoolbox}

\pretocmd{\thebibliography}{%
  \sloppy
  \emergencystretch=1em
}{}{}

\title{Gravitational Wave Signatures from Lepton Number Breaking Phase Transitions with Flat Potentials}

\author[a,b]{Gabriela Barenboim}
\emailAdd{gabriela.barenboim@uv.es}

\author[c]{Yeji Park,}

\emailAdd{yejipark@ibs.re.kr}

\author[d,e]{Liliana Velasco-Sevilla}

\emailAdd{lilianak@sogang.ac.kr}
\affiliation[a]{Instituto de F\'{i}sica Corpuscular, CSIC-Universitat de Val\`{e}ncia, Paterna 46980, Spain}
\affiliation[b]{Departament de F\'{i}sica Te\`{o}rica, Universitat de Val\`{e}ncia, Burjassot 46100, Spain}

\affiliation[c]{Particle Theory and Cosmology Group, Center for Theoretical Physics of the Universe, Institute for Basic Science (IBS), Daejeon 34126, Republic of Korea}

\affiliation[d]{Center for Quantum Spacetime, Sogang University, 35 Baekbeom-ro, 
\\Seoul 121-742, South Korea}

\affiliation[e]{Department of Physics, Sogang University, 
\\35 Baekbeom-ro, Seoul 121-742, South Korea}

\abstract{Extensions of the Standard Model typically contain ``flaton fields" defined as fields with large vacuum expectation values and almost flat potentials where scalar self-coupling is small or vanishes at tree level. Such potentials have been used to drive a secondary inflationary epoch after a primary phase of inflation, in what are called thermal inflation models. Although the primordial, high-scale inflationary epoch can solve the horizon and flatness problems, it does not always resolve difficulties associated with late-time relics produced in extensions of the Standard Model. These relics typically decay too late, injecting entropy 
and energetic particles that spoil successful predictions like Big Bang Nucleosynthesis. It is here that thermal inflation plays a crucial role: diluting unwanted relics by many orders of magnitude without 
erasing the baryon asymmetry or the large-scale structure set up by the 
earlier phase of inflation.  The preferred scale for this phenomenon is in the range $10^6-10^8$  GeV if one considers supergravity, but without it, any scale above the EW scale is valid. We investigate a typical form of these potentials and determine what are the conditions for the potentials to develop a barrier such that when the flatons settle to the true minimum, the associated Gravitational Waves can be observed,
focusing on first-order phase transitions from spontaneous lepton number breaking.
}

\begin{document}

\begin{flushright}
CQUeST-2026-0769
\end{flushright}
\maketitle
\flushbottom

\section{Introduction}

The detection of gravitational waves (GWs) has opened a new observational window into the physics of the early universe, enabling the study of phenomena that are otherwise inaccessible to direct experimental probes. 

In particular, stochastic GW backgrounds sourced by cosmological first-order phase transitions (FOPTs) provide a unique means of exploring high-energy particle physics beyond the Standard Model (SM). Such transitions are predicted in a wide variety of extensions to the SM, including models with extended scalar sectors, hidden gauge groups, and additional global symmetries. In the presence of a FOPT, the dynamics of bubble nucleation, expansion, and collision, together with the resulting sound waves and magnetohydrodynamic turbulence in the plasma, can lead to the production of a GW spectrum whose characteristics are directly tied to the underlying microphysics.

Among the most theoretically compelling settings for such a transition is the spontaneous breaking of lepton number symmetry, which dynamically generates Majorana masses for right-handed neutrinos. The observation of neutrino oscillations~\cite{Fukuda:1998mi,Abe:2013tm} has firmly established that neutrinos have nonzero masses and mix~\cite{Bilenky:1998dt,GonzalezGarcia:2002dz,deSalas:2018bym,deSalas:2020pgw}, a fact that cannot be accommodated within the SM without introducing new physics. The seesaw mechanism offers an elegant explanation for these masses, naturally linking them with lepton-number-violating physics at high energy scales \footnote{~\cite{Minkowski:1977sc}, for the Minkowski’s original Type I framework, the formulations by Yanagida~\cite{Yanagida:1979as} and by Gell-Mann, Ramond, and Slansky~\cite{GellMann:1979vx}, and the left–right symmetric / Type II versions developed by Mohapatra and Senjanović~\cite{Mohapatra:1980yp}. }. Moreover, the same framework can account for the observed matter–antimatter asymmetry of the universe via leptogenesis~\cite{Fukugita:1986hr}, providing a unified picture of neutrino mass generation and baryogenesis \cite{Addazi:2023majorsignals, Addazi:2019footprints, DiBari:2021majoronGW, Madge:2018leptophilic, Bosch:2023gauged,Jana:2025vyb}.

When lepton number symmetry breaking occurs through a FOPT in the early universe, it can produce a stochastic GW background with a characteristic peak frequency determined by the nucleation temperature and the inverse duration parameter $\beta/H$, and an amplitude governed by the transition strength parameter $\alpha$. The precise values of these parameters depend sensitively on the scalar potential, the vacuum expectation value associated with the symmetry breaking, and the thermal corrections from the particle content. 
Lepton number symmetry breaking can come from a step in the chain of a Grand Unified Theory breaking down to the SM. What we propose here is that the potential associated with it is flat and can lead to thermal inflation \cite{Lyth:1995ka,Lyth:1995hj}. This can happen when some heavy fields get integrated out leaving a flat potential. We give a simple example of this in Appendix \ref{app:heavyfields}. In these flat potential scenarios, thermal corrections become the dominant source of barrier formation between symmetric and broken phases, typically leading to stronger first-order transitions and enhanced gravitational wave signals. In thermal inflation scenarios, the cosmological history can significantly alter these quantities, shifting the resulting GW spectrum into the sensitivity ranges of planned space-based interferometers.


This work provides the first systematic study connecting flat scalar potentials to neutrino mass generation via the seesaw mechanism, with a detailed analysis of symmetry-breaking patterns in both $\mathrm{U}(1)$ and $\mathrm{SU}(2)$ frameworks. We investigate the theoretical and phenomenological implications of the associated high-scale cosmological phase transition, emphasizing its detectability in forthcoming gravitational-wave observatories such as LISA~\cite{LISA:2017pwj}, DECIGO~\cite{Kawamura:2011zz}, and BBO~\cite{Crowder:2005nr}. A striking outcome of the analysis is the emergence of a nontrivial scaling of the potential barrier, $V \propto m_0^{3} m_X$, which persists despite strong $1/m_X^{2}$ suppression of the sextic operator, revealing a subtle and unexpected enhancement mechanism. Observation of the predicted signal would constitute direct evidence for a high-scale phase transition, indirect confirmation of lepton-number violation in the neutrino sector, and strong support for the seesaw origin of neutrino masses, while null results would still impose powerful constraints on the parameter space of lepton-number-breaking models, underscoring the rich dynamics at the interface of neutrino physics and scalar field theory.

In the following sections, we present a detailed study of the finite-temperature effective potential governing the symmetry-breaking sector, Section \ref{sec:efflatPot}, calculate the resulting GW spectrum, and explore the experimental prospects for observing such a signal, Section \ref{sec:ResFOPT}. By connecting early-universe cosmology to the properties of the neutrino sector, this work illustrates the potential of gravitational wave astronomy to probe fundamental questions about mass generation, symmetry breaking, and the thermal history of the universe, we discuss this in Section \ref{sec:Discussion}.

\section{Effective Flat Potential at Finite Temperature \label{sec:efflatPot}}

The dynamics of a gauge theory breaking phase transition are determined by the finite-temperature effective potential, which incorporates both quantum and thermal corrections to the tree-level potential. For a complex scalar field $\Phi$ responsible for the symmetry breaking, the general one-loop, plus daisy resummation term included, reads
\begin{equation}
V_{\mathrm{Eff. }}(\Phi, T) = V_{\mathrm{Tree}}(\Phi) + V_{\mathrm{CW}}(\Phi) + V_{T}(\Phi, T) + V_{\mathrm{Daisy}}(\Phi, T),
\label{eq:Veff}
\end{equation}
where we work in terms of the field magnitude $|\Phi|$ for rotationally symmetric potentials. In the case of the quartic coupling scalar potential, the tree-level potential is $V_{\mathrm{Tree}}(\Phi) = -\frac{\mu^2}{2}|\Phi|^2 + \frac{\lambda}{4}|\Phi|^4 $, with $\mu^2>0$ to ensure spontaneous symmetry breaking at $T=0$ and the well known form of the Coleman--Weinberg zero-temperature \cite{Coleman:1973jx} one-loop correction \footnote{$V_{\mathrm{CW}}(\Phi) = \sum_{i} \frac{n_i}{64\pi^2} m_i^4(\Phi)
    \left[ \ln\!\left(\frac{m_i^2(\Phi)}{\mu_R^2}\right) - c_i \right]$, where the sum runs over all particle species $i$ coupled to $\Phi$, $n_i$ counts the degrees of freedom (positive for bosons, negative for fermions), $m_i(\Phi)$ are the field-dependent masses, $\mu_R$ is the renormalisation scale, and $c_i$ are scheme-dependent constants ($3/2$ for scalars/fermions, $5/6$ for gauge bosons in $\overline{\text{MS}}$.} \cite{PhysRevD.9.3320,Linde:1981zj}.  $V_{T}(\Phi, T)$ is the one-loop finite-temperature correction,
    \begin{equation}
    V_{T}(\Phi, T) = \frac{T^4}{2\pi^2} \left[ 
        \sum_{i \in \mathrm{bosons}} n_i J_B\!\left( \frac{m_i^2(\Phi)}{T^2} \right) +
        \sum_{i \in \mathrm{fermions}} n_i J_F\!\left( \frac{m_i^2(\Phi)}{T^2} \right)
    \right],
    \label{eq:VT}
    \end{equation}
    where $J_{B,F}$ are the standard bosonic and fermionic thermal functions, \eq{eq:JBJF}.
Finally, $V_{\mathrm{Daisy}}(\Phi, T)$ incorporates ring (Daisy) resummations for bosonic zero modes, achieved by the following replacement:
    \begin{equation}
    m_i^2(\Phi) \;\rightarrow\; m_i^2(\Phi) + \Pi_i(T).
    \end{equation}
This procedure then gives the form of $V_{\mathrm{Daisy}}(\Phi, T)$ of \eq{eq:VDaisy}, in the
    high-temperature expansion of $J_B$, with $\Pi_i(T)$ representing the thermal self-energy corrections that account for the effective mass acquired by particles in the hot plasma environment - these corrections are essential to avoid infrared divergences in the high-temperature limit.

In the high-temperature limit ($y=m_i/T \ll 1$), the thermal functions admit the familiar expansions of \eq{eq:JBJF_exp}.
 The cubic bosonic term $\propto -y^3$ is responsible for generating a barrier between the symmetric and broken phases, enabling a first-order transition.

For quartic potentials, thermal corrections can alter the shape of $V_{\mathrm{Eff. }}$, either weakening or strengthening the transition depending on the couplings and particle content. In ``flat'' scenarios, where $\lambda$ is small or vanishes at tree level, thermal effects may dominate the potential, leading to a stronger first-order transition and a potentially enhanced gravitational wave signal.

All explicit formulas for $V_{T}$, $\Pi_i(T)$, and the numerical treatment of thermal integrals are provided in Appendix~\ref{app:furthercontThPot}, while the details of the FOPT and GW parameters are given in Appendix~\ref{ap:FOPTandGW_nt}.

In the next section, we apply this finite-temperature formalism to the extended scalar sectors of our model, identifying the relevant particle spectrum, computing the temperature-dependent masses, and evaluating the resulting phase-transition parameters. 
We focus on $U(1)$ and $SU(2)$ extensions since they are the simplest cases that capture qualitatively distinct dynamics. The SU(2) case is also particularly relevant for neutrino mass models via seesaw-type interactions.

\subsection{$U(1)$ case \label{sec:u1}}

We begin our analysis with an effective $U(1)$ gauge model described by the Lagrangian
\bea
\mathcal{L} &=& (D^\mu \Phi)^\dagger (D_\mu \Phi) - V(\Phi) -\frac{1}{4} F^{\mu\nu}F_{\mu\nu} \nonumber \\
&+& \sum_{i=1}^{n^D_f} \left[ \overline{\psi}_{Li} i \gamma^\mu D_\mu \psi_{Li} + \overline{\psi}_{Ri} i \gamma^\mu D_\mu \psi^c_{Ri} - \frac{y}{2} \left( \overline{\psi}_{Li} \Phi \psi^c_{Li} + \overline{\psi}_{Ri} \Phi \psi^c_{Ri} + \text{h.c.} \right) \right] \nonumber \\
&-& \sum_{i=1}^{n^M_f}\left[ y_f \Phi \overline{\psi^c_M} \psi_M +\text{h.c.} \right],
\eea
where $\Phi$ is a complex scalar field charged under the $U(1)$ gauge symmetry, and the covariant derivative is defined as $D_\mu = \partial_\mu - i g A_\mu$, with $g$ the gauge coupling and $A_\mu$ the gauge field. The fields $\psi_L$ and $\psi_R$ represent the left- and right-handed components of Dirac fermions, with $n^D_f$ fermion flavors. The Yukawa coupling $y$ generates fermion masses after symmetry breaking and is allowed only if the $U(1)$ charges satisfy the charge conservation condition
\bea
c_{\overline{\psi}_{Li}} + c_{\Phi} + c_{\psi^c_{Li}} = 0,
\eea
where $c$ denotes the $U(1)$ charge of the corresponding field. For example, if we normalize the charge of $\Phi$ to be unity, then $\psi$ has charge $1/2$ so that Yukawa couplings are permitted. We have also written the Majorana mass term for the Majorana fermion $\psi_M$. 
An effective mass term with mass parameter $m$ may also arise. The classical effective potential for the $U(1)$ scalar field $\Phi$ can be written as \cite{Barreiro:1996dx}
\begin{align}
V_\text{Tree}(\Phi) = V_0 - m_0^2 \Phi^\dagger \Phi + \frac{\lambda_n}{m_X^{2n}} \left(\Phi^\dagger \Phi \right)^{n+2},
\label{eq:general_Lag_not}
\end{align}
After spontaneous symmetry breaking, we can parameterize the complex field as $\Phi = \frac{1}{\sqrt{2}}(\phi + i\chi)$ where $\phi$ is the real radial mode that acquires a vacuum expectation value $v = \langle\phi\rangle$, and $\chi$ is the Goldstone boson mode. The potential expressed in terms of the radial field $\phi$ becomes
\bea
\Phi = \frac{v + \delta \phi + i \chi}{\sqrt{2}},
\eea
where $\delta \phi$ and $\chi$ represent quantum fluctuations around the classical vacuum value. Specifically, $\delta \phi$ corresponds to the real, massive mode (radial field), while $\chi$ represents the Goldstone boson resulting from the broken symmetry. Here, $m_0$ sets the low-energy mass scale, $m_X$ denotes the UV cutoff or maximum scale (often associated with the Planck scale $\Mp = 1.2 \times 10^{19}$ GeV\footnote{Note that we follow the normalization conventions of \cite{Barreiro:1996dx} such that the maximal scale is the reduced Planck scale $\overline{M}_{\rm Pl} = \Mp / \sqrt{8 \pi}$.}), and $\lambda_n$ is the self-coupling parameter of the scalar field.

Expressing the potential explicitly in terms of the classical scalar field magnitude $\phi = \sqrt{2}|\Phi|$, we have
\begin{align}
V_{\mathrm{Tree}}(\phi) = V_0 - \frac{1}{2} m_0^2 \phi^2 + \frac{\lambda_n}{m_X^{2n}} \left( \frac{\phi^2}{2} \right)^{n+2},
\label{eq:general_Lag}
\end{align}
The vacuum expectation value (vev) $v = \langle\phi\rangle$, the scalar mass $m_\phi$ \footnote{That is
$m_\phi^2(\phi) = \frac{\partial^2 V}{\partial\phi^2}\bigg|_{\phi}.$},  and the vacuum energy $V_0$ relate to $m_0$, $m_X$, and $\lambda_n$ as
\bea
\label{eq:massesmphi}
m_\phi^2 &=& 2 (n+1) m_0^2, \nonumber \\
v^{2(n+1)} &=& \frac{1}{\lambda_n} \frac{1}{n+2} m_X^{2n} m_0^2, \nonumber \\
V_0 &=& \frac{n+1}{2 (n+2)} v^2 m_0^2,
\eea
where these relations ensure the potential is minimized at $\phi = v$. In this work, we focus on the case $n=1$, because it gives the leading sextic term after higher-dimensional operators, and it is the simplest case capturing essential physics. The case $n=2$ corresponds to the power of eight, and its effects are suppressed in comparison to the sextic term effects.  Minimizing the potential for $n=1$ yields field-dependent masses for the scalar (radial mode) field and its associated Goldstone boson:
\begin{align}
m_\phi^2(\phi) &= \frac{15}{4} \frac{\lambda}{m_X^2} \phi^4 - m_0^2, \\
m_\chi^2(\phi) &= \frac{3}{4} \frac{\lambda}{m_X^2} \phi^4 - m_0^2,
\end{align}
where $m_\phi(\phi)$ is the mass of the radial mode, and $m_\chi(\phi)$ is the mass of the Goldstone boson. The different coefficients (15/4 vs 3/4) arise from the curvature of the potential in the radial and angular directions, respectively. We assume that the fermion mass parameter $m$ is much smaller than the vev parameter $v$. The gauge boson and fermion masses that couple to the scalar depend on the field value $\phi$ as
\begin{align}
m_b^2(\phi) &= g^2 \phi^2, \\
m_f^2(\phi) &= \frac{y^2}{2} \phi^2,
\end{align}
where $m_b(\phi)$ is the gauge boson mass associated with the $U(1)$ symmetry, and $m_f(\phi)$ corresponds to fermions interacting with the scalar. Here $g$ and $y$ denote the gauge and Yukawa couplings, respectively. For simplicity, the benchmark points presented below omit fermions, as their inclusion leads to only mild quantitative differences in the results.
\begin{figure}[h]
    \centering
    \includegraphics[width=0.495\textwidth]{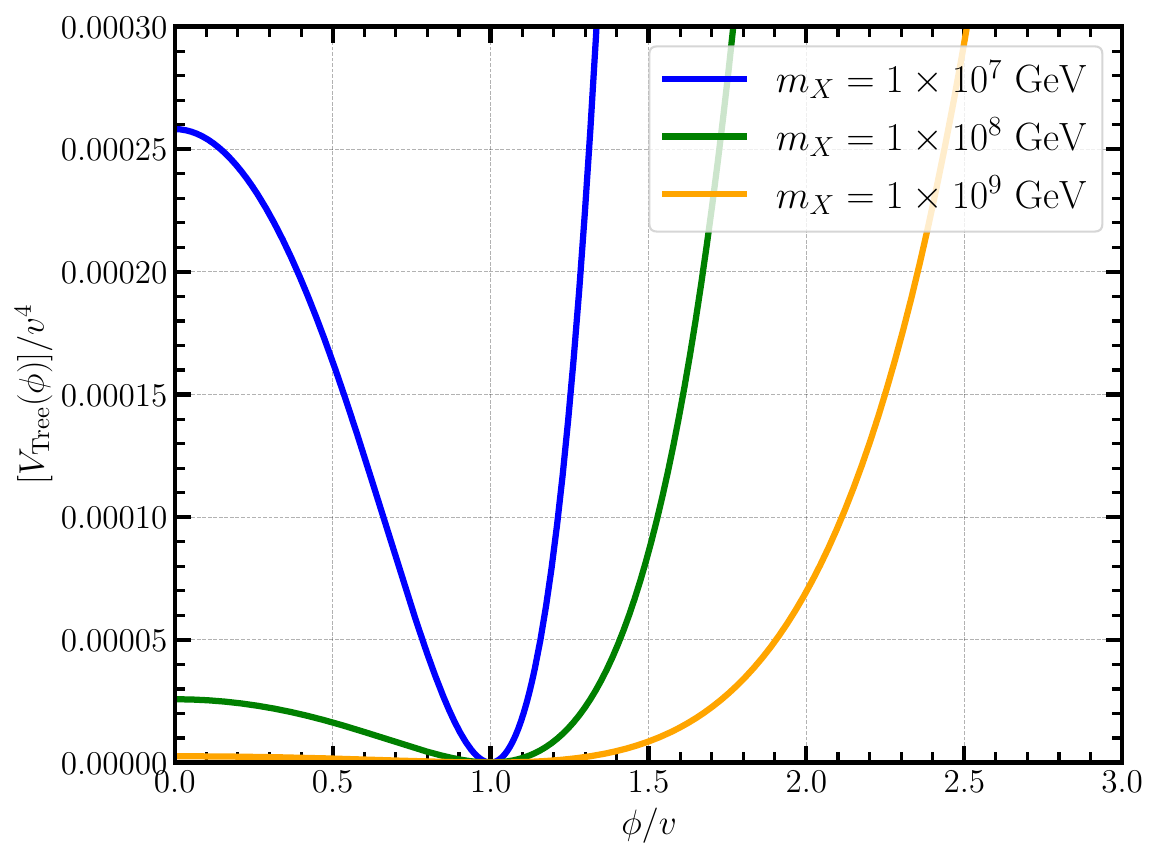}
    \includegraphics[width=0.495\textwidth]{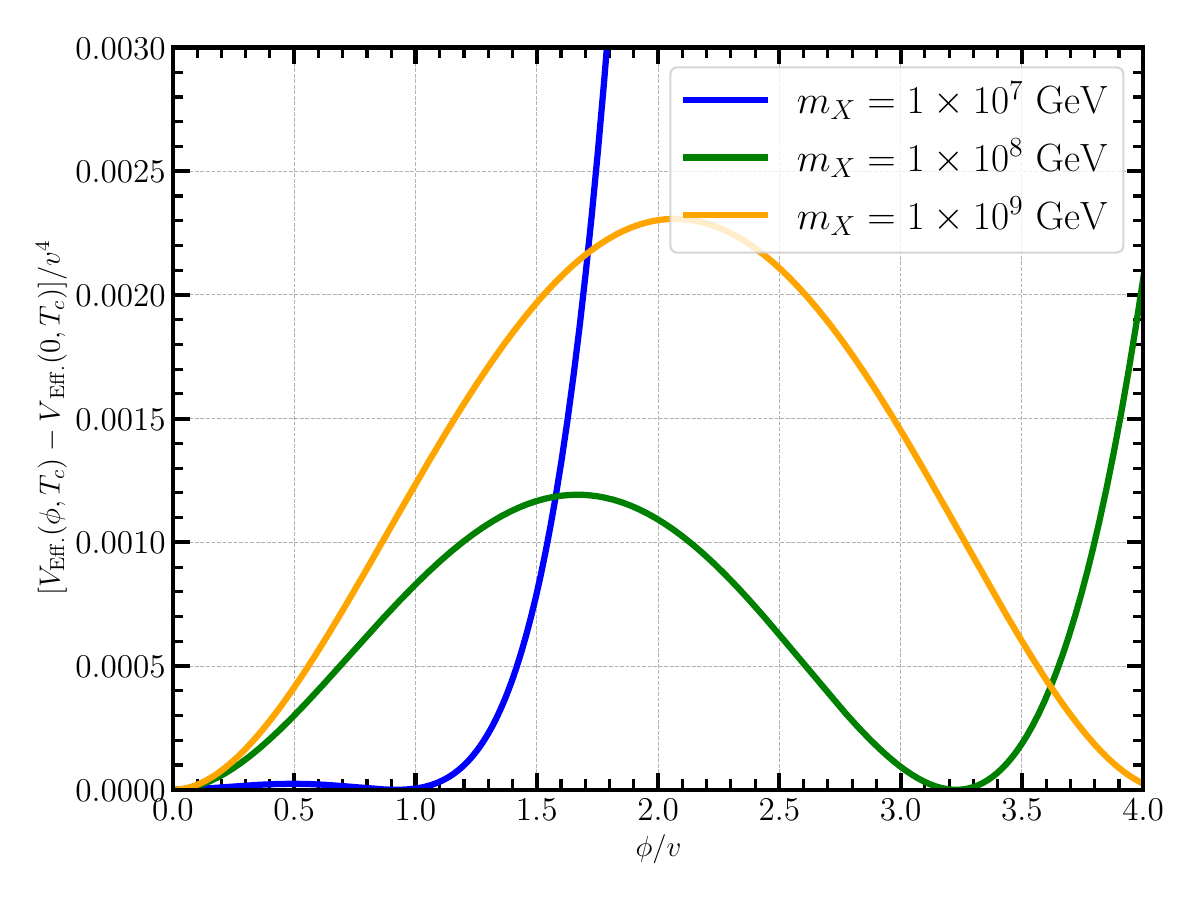}
    \caption{U(1) symmetry breaking potentials for various $m_X$ values with coupling parameters $g=0.3$ and $\lambda=0.8$, for a fixed value of $m_0=10^4$ GeV. \textbf{Left:} tree-level potential $V(\phi)/v^4$ versus $\phi/v$ for $m_X = 10^7$, $10^8$, and $10^9$ GeV,  illustrating how the potential flattens as $m_X$ increases. \textbf{Right:} Total effective potential at the critical temperature $T_c$, demonstrating that larger $m_X$ values produce higher and wider potential barriers. Here, $v$ denotes the vacuum expectation value of the tree-level potential. The two plots have been rendered in different scales, due to the size of the loop and thermal contributions.}
    \label{fig:U(1)_potentials}
\end{figure}

\begin{table}
\centering
\renewcommand{\arraystretch}{1.5}
\begin{tabular}{|c|c|c|c|c|c|c|c|c|}
\toprule
\hline
\multicolumn{3}{|c|}{\textbf{Model parameters}} & \textbf{VEV} & \multicolumn{3}{c|}{\textbf{Temperatures}} & \multicolumn{2}{c|}{\textbf{Phase Trans.}} \\ \hline
$m_X$ & $g$ & $\lambda$ & $v$ & $T_0$ & $T_n$ & $T_c$ & $\alpha_n$ & $\beta/H_n$ \\ \hline
 [GeV] &  &  & [GeV] & [GeV] & [GeV] & [GeV] & &  \\ \hline \hline
$1.0 \times 10^{7}$ & $0.3$ & $0.8$ & $3.6 \times 10^{5}$ & $6.7 \times 10^{4}$ & $7.0 \times 10^{4}$ & $9.2 \times 10^{4}$ & $5.2 \times 10^{-3}$ & $3.24 \times 10^{3}$ \\ \hline
$1.0 \times 10^{7}$ & $0.3$ & $0.9$ & $3.5 \times 10^{5}$ & $6.7 \times 10^{4}$ & $7.0 \times 10^{4}$ & $9.0 \times 10^{4}$ & $4.7 \times 10^{-3}$ & $3.26 \times 10^{3}$ \\ \hline
$1.0 \times 10^{7}$ & $0.4$ & $0.8$ & $3.6 \times 10^{5}$ & $5.0 \times 10^{4}$ & $5.5 \times 10^{4}$ & $9.4 \times 10^{4}$ & $3.3 \times 10^{-2}$ & $1.69 \times 10^{3}$ \\ \hline
$1.0 \times 10^{7}$ & $0.4$ & $0.9$ & $3.5 \times 10^{5}$ & $5.0 \times 10^{4}$ & $5.5 \times 10^{4}$ & $9.2 \times 10^{4}$ & $2.9 \times 10^{-2}$ & $1.69 \times 10^{3}$ \\ \hline
\hline
$1.0 \times 10^{8}$ & $0.3$ & $0.8$ & $1.1 \times 10^{6}$ & $6.7 \times 10^{4}$ & $7.0 \times 10^{4}$ & $2.6 \times 10^{5}$ & $9.0 \times 10^{-1}$ & $2.87 \times 10^{3}$ \\ \hline
$1.0 \times 10^{8}$ & $0.3$ & $0.9$ & $1.1 \times 10^{6}$ & $6.7 \times 10^{4}$ & $7.1 \times 10^{4}$ & $2.5 \times 10^{5}$ & $7.5 \times 10^{-1}$ & $3.02 \times 10^{3}$ \\ \hline
$1.0 \times 10^{8}$ & $0.4$ & $0.8$ & $1.1 \times 10^{6}$ & $5.0 \times 10^{4}$ & $5.6 \times 10^{4}$ & $3.4 \times 10^{5}$ & $8.0 $ & $1.37 \times 10^{3}$ \\ \hline
$1.0 \times 10^{8}$ & $0.4$ & $0.9$ & $1.1 \times 10^{6}$ & $5.0 \times 10^{4}$ & $5.6 \times 10^{4}$ & $3.3 \times 10^{5}$ & $7.0 $ & $1.39 \times 10^{3}$ \\ \hline
\hline
$1.0 \times 10^{9}$ & $0.3$ & $0.8$ & $3.6 \times 10^{6}$ & $6.7 \times 10^{4}$ & $7.0 \times 10^{4}$ & $1.1 \times 10^{6}$ & $4.3 \times 10^{2}$ & $2.82 \times 10^{3}$ \\ \hline
$1.0 \times 10^{9}$ & $0.3$ & $0.9$ & $3.5 \times 10^{6}$ & $6.7 \times 10^{4}$ & $7.0 \times 10^{4}$ & $1.1 \times 10^{6}$ & $3.7 \times 10^{2}$ & $2.39 \times 10^{3}$ \\ \hline
$1.0 \times 10^{9}$ & $0.4$ & $0.8$ & $3.6 \times 10^{6}$ & $5.0 \times 10^{4}$ & $5.6 \times 10^{4}$ & $1.3 \times 10^{6}$ & $1.7 \times 10^{3}$ & $1.09 \times 10^{3}$ \\ \hline
$1.0 \times 10^{9}$ & $0.4$ & $0.9$ & $3.5 \times 10^{6}$ & $5.0 \times 10^{4}$ & $5.3 \times 10^{4}$ & $1.2 \times 10^{6}$ & $1.8 \times 10^{3}$ & $1.05 \times 10^{3}$ \\ \hline
\end{tabular}
\caption{Benchmark points for the $U(1)$ symmetry breaking scenario without fermions ($n_f=0$) for a fixed mass parameter $m_0 = 10^4\,\mathrm{GeV}$. Results show the evolution of transition strength parameter $\alpha_n$ with UV cutoff scale $m_X$ for various values of gauge coupling $g$ and scalar self-coupling $\lambda$. The thermal corrections to particle masses are included as described in Appendix \ref{app:furthercontThPot}.}
\label{tab:U(1)_params}
\end{table}
Since we are working in the regime where the contributions depending on the field $\phi$ dominate over temperature-dependent effects, we take into account the finite-temperature self-energy diagrams of \fig{fig:Finite_Tcorr}, and we resummate them, following \cite{Carrington:1991hz, Arnold:1992rz}.
 
In \fig{fig:grid_GWU1} we present the shape of the tree-level potential (left panel) and the thermally corrected (right panel) potential, where we can clearly see the thermal origin of the barrier for different scales, $m_X=10^7, 10^8, 10^9$ GeV, for $m_0 = 10^4$ GeV and $n_f=0$. This choice is motivated by collider constraints, which allow for new physics to emerge at scales starting from $m_0 = 10^4$ GeV. Given this lower bound, we explore what scale is required for a first-order phase transition (FOPT) associated with lepton symmetry breaking to occur. Our analysis shows that such a transition can begin to appear at around $10^7$ GeV. In \tab{tab:U(1)_params} we present some benchmark points.  For the thermally corrected potential we have used \eq{eq:Veff} with \eq{eq:VT}, using the complete numerical form of the thermal functions $J_B$ and $J_F$, \eq{eq:JBJF}, and the thermal energy corrections given in Appendix \ref{app:furthercontThPot}. We have employed our own code based on \texttt{CosmoTransitions} \cite{Wainwright:2011kj} functions.

For the breaking of $U(1)$, a well--known consequence of Abelian symmetry breaking is the generation of cosmic strings. In our case, such strings do not lead to any observational issues since their tension is far below current constraints. The CMB bound requires
$G\mu < 10^{-7}$, while for the relevant range of symmetry--breaking scales considered in this work, $v \sim 10^{6}\!-\!10^{9}\,\text{GeV}$, the resulting tensions are  $G\mu \sim 10^{-26}\!-\!10^{-20}$. These signals are too faint to be seen by current and proposed experiments in the frequencies that are the interest of this work (above $10^{-5}$ Hz) and see Figs.~\ref{fig:GW_densityU1}-\ref{fig:GW_densitySU2not}.

\subsection{$SU(2)$ case}
In contrast to the abelian $U(1)$ case discussed previously, the $SU(2)$ gauge symmetry introduces a richer structure due to its non-abelian nature. 
First, the scalar field can transform either in the adjoint or fundamental representation, leading to different patterns of symmetry breaking ($SU(2)\to U(1)$ or $SU(2)\to 1$). 
Second, the number of gauge boson degrees of freedom is larger, which amplifies thermal loop effects and strengthens the resulting potential barrier. 
Finally, the possible fermion representations and Yukawa couplings differ from the $U(1)$ case, with right-handed neutrinos in particular playing a more direct role in mass generation. 
We now present the formal setup for the $SU(2)$ case.

The starting point is the Lagrangian
\bea
\mathcal{L} = (D^\mu \Phi)^\dagger (D_\mu \Phi) - V(\Phi),
\eea
where $\Phi$ is a complex scalar field transforming under $SU(2)$, and the covariant derivative is defined as
\bea
D_\mu \Phi_a = \partial_\mu \Phi_a - i g A^b_{\mu a} \Phi_b,
\eea
with $a,b = 1,2$ labeling the $SU(2)$ indices and $g$ the gauge coupling constant. If fermions are introduced, at least two different representations are required to write Yukawa couplings invariant under $SU(2)$ – for example, one doublet and one singlet representation – so that couplings between the fermions and the scalar field are allowed. The well-known generators of $SU(2)$ are the Pauli matrices, and we choose the conventional form of generators $\tau_i$ as follows:
\bea
\label{eq:paulimatrices}
\sigma_1 = \begin{pmatrix} 0 & 1 \\ 1 & 0 \end{pmatrix}, \quad
\sigma_2 = \begin{pmatrix} 0 & -i \\ i & 0 \end{pmatrix}, \quad
\sigma_3 = \begin{pmatrix} 1 & 0 \\ 0 & -1 \end{pmatrix},\quad \tau_i = \frac{1}{2}\sigma_i.
\eea
There are two possible patterns of symmetry breaking:
\bea
SU(2) \longrightarrow U(1),
\eea
and
\bea
SU(2) \longrightarrow \mathds{1},
\eea
the first one achieved through the adjoint representation of $SU(2)$ while the second through the fundamental. We discuss both of them in the following sections.

\subsection{$SU(2) \longrightarrow U(1)$ \label{subsec:su2tu1}}

In this breaking pattern, the adjoint representation allows the scalar field to develop a vacuum expectation value that preserves a $U(1)$ subgroup of $SU(2)$, corresponding to rotations around a specific axis in the internal space. This partial breaking is physically relevant in models where gauge symmetry is broken in stages, potentially connecting to electroweak or grand unified theories \footnote{In this $SU(2)\to U(1)$ breaking, monopoles arise since the homotopy groups
$\pi_2(SU(2)/U(1))\cong\pi_1(U(1))\cong\mathbb{Z}$. 
To remove them, one can embed the theory in a larger simply connected group
$\widetilde{G}$ broken to a subgroup $\widetilde{H}$ with trivial $\pi_1(\widetilde{H})$. 
For example, $SU(3)\to SU(2)$ yields $\widetilde{G}/\widetilde{H}\cong S^5$ with
$\pi_1=\pi_2=\mathds{1}$, so no monopoles or strings appear. The embedding is natural in this context because we are thinking that this breaking may be happening at a given stage in the breaking route from the GUT group to the SM. Another possibility is that the brief period of thermal inflation dilutes the monopole abundance through entropy production, leading to an exponential suppression of the yield, with the final abundance set by the reheating temperature.}. In this scenario, the Lagrangian takes the following form:
\begin{align}
   \mathcal{L} = -\frac{1}{4} F_{\mu\nu}^a F^{a\,\mu\nu} + \frac{1}{2} (D_\mu \Phi)^\mathrm{T} (D^\mu \Phi) - V(\Phi).
\end{align}
For the adjoint representation, we define the scalar field as
\bea
\Phi = \phi_i \tau_i, \quad i = 1,2,3,
\eea
where we expand the scalar field $\Phi$ around the vev as follows
\bea
\label{eq:sola}
\Phi= \begin{pmatrix} 0 \\ 0 \\ \phi_3+ v \end{pmatrix}.
\eea
This vacuum configuration also satisfies
\bea
\left[ \tau^i, \langle \Phi \rangle \right] = 0,
\eea
for $i=3$, meaning the vacuum is invariant under the $U(1)$ subgroup generated by $\tau_3$. The unbroken $U(1)$ generator $\tau_3$ commutes with the VEV, while the broken generators $\tau_1$ and $\tau_2$ do not, leading to two massive gauge bosons and one massless gauge boson. Using the $SU(2)$ algebra
\[
[\tau_i, \tau_j] = i \epsilon_{ijk} \tau_k,
\]
we see that the commutator vanishes for this choice since the vev points in the $i=3$ direction, commuting with $\tau_3$. By cyclic permutations of indices, analogous vacua exist with vevs aligned along $i=1$ or $i=2$ directions.From now on we drop the index 3 in \eq{eq:sola}. Before symmetry breaking, the tree-level potential is written as
\begin{align}
    V_\text{Tree}(\Phi) = V_0 - \frac{m_0^2}{2} \Phi^2 + \frac{1}{8} \frac{\lambda_n}{m_X^{2n}} \Phi^{2(n+2)},
    \label{eq:su2V}
\end{align}
where $m_0$ and $\lambda_n$ are parameters as defined in the $U(1)$ case.
After alignment, the potential takes the same functional form as in \eq{eq:general_Lag} and the Eqs.~ (\ref{eq:massesmphi}) apply for this case.
 In this breaking pattern, $SU(2)$ is broken down to $U(1)$, leaving one unbroken gauge generator and therefore two massive gauge bosons, associated with the broken generators, with mass
\begin{equation}
m_b^2(\phi) = g^2 \phi^2.
\end{equation}
\begin{figure}[h]
\begin{subfigure}[b]{0.49\textwidth}
    \includegraphics[width=\linewidth]{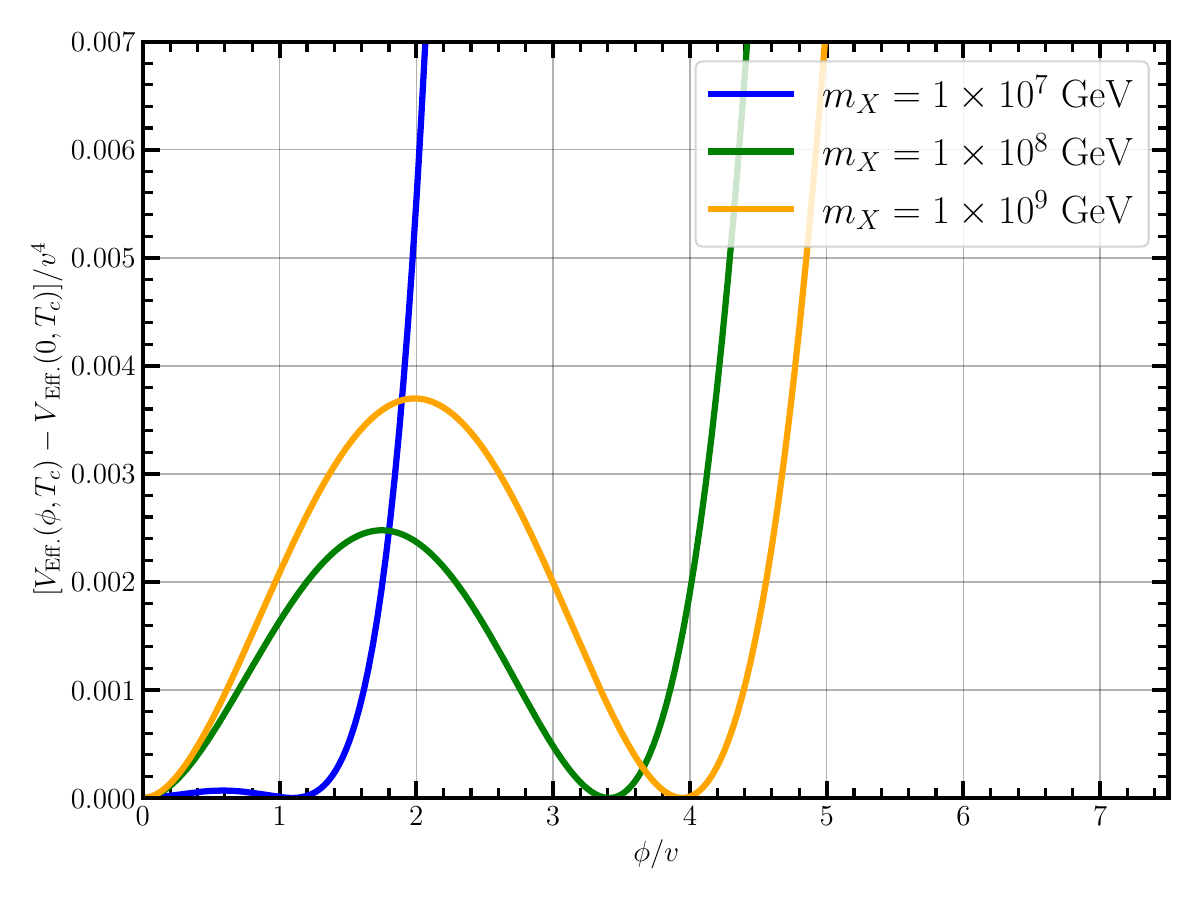}
    \caption{$SU(2)\rightarrow U(1)$.}
 \end{subfigure}
\begin{subfigure}[b]{0.49\textwidth} 
    \includegraphics[width=\linewidth]{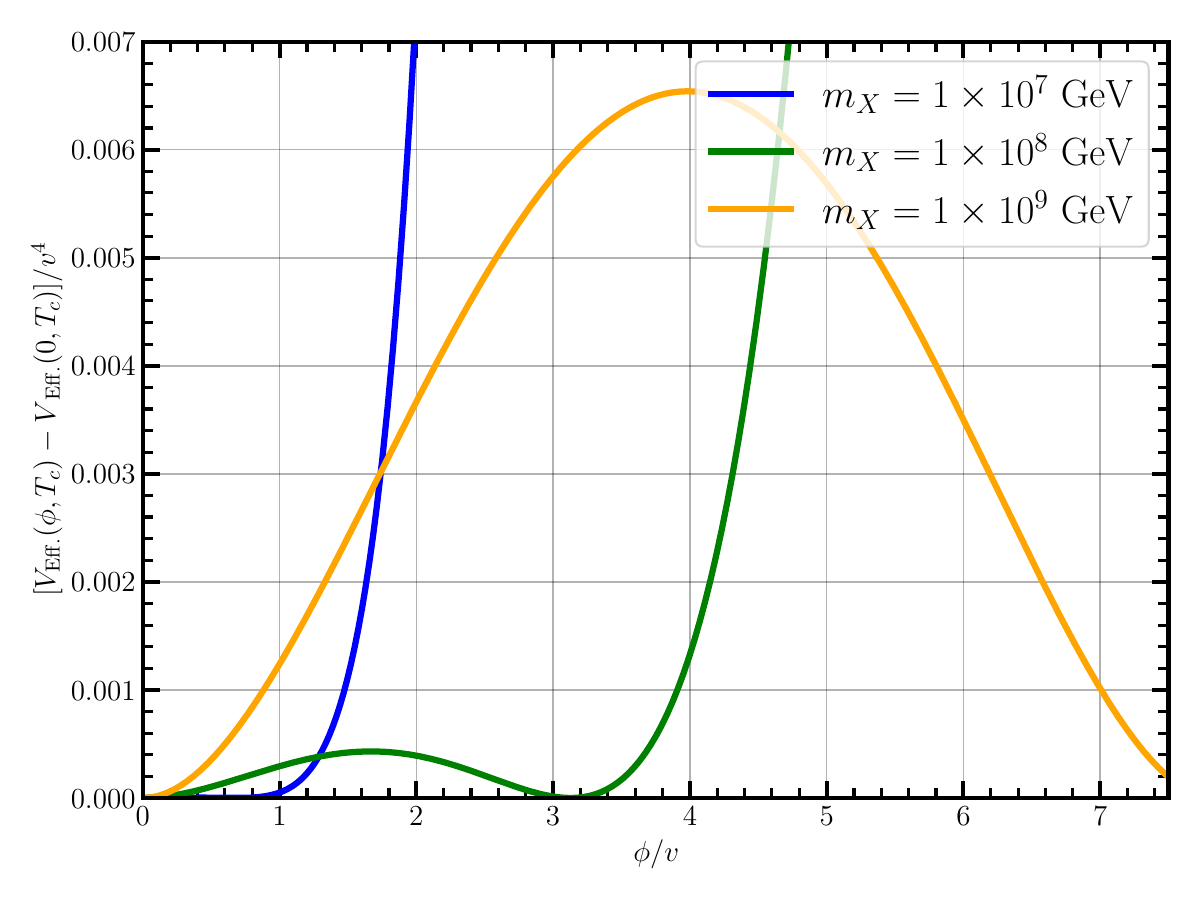}
 \caption{$SU(2)\rightarrow \mathds{1}$.}
 \end{subfigure}
    \caption{{\bf Left}: $SU(2)$ symmetry partially breaking potentials for different $m_X$ with coupling parameters $g=0.3$, $m_0=10^4$ GeV, and $\lambda=0.8$ and $n_f=0$. Total effective potential at critical temperature $T_c$, showing how increasing $m_X$ leads to higher and wider potential barriers. {\bf Right}: $SU(2)$ symmetry maximally breaking potential for different $m_X$ with coupling parameters $g=0.3$ and $\lambda=0.8$, for this case also $n_f=0$. The effective thermal potential is shown at the critical temperature $T_c$, showing how increasing $m_X$ leads to higher and wider potential barriers. In both cases, $v$ is the vacuum expectation value of the tree-level potential.
    \label{fig:SU(2)_all}}
\end{figure}
\begin{table}
\centering
\renewcommand{\arraystretch}{1.5}
\begin{tabular}{|c|c|c|c|c|c|c|c|c|}
\toprule
\hline
\multicolumn{3}{|c|}{\textbf{Model parameters}} & \textbf{VEV} & \multicolumn{3}{c|}{\textbf{Temperatures}} & \multicolumn{2}{c|}{\textbf{Phase Trans.}} \\ \hline
$m_X$ & $g$ & $\lambda$ & $v$ & $T_0$ & $T_n$ & $T_c$ & $\alpha_n$ & $\beta/H_n$ \\ \hline
 [GeV] &  &  & [GeV] & [GeV] & [GeV] & [GeV] & &  \\ \hline \hline
$1.0 \times 10^{7}$ & $0.3$ & $0.8$ & $3.6 \times 10^{5}$ & $5.4 \times 10^{4}$ & $5.5 \times 10^{4}$ & $8.0 \times 10^{4}$ & $2.5 \times 10^{-2}$ & $4.59 \times 10^{3}$ \\ \hline
$1.0 \times 10^{7}$ & $0.3$ & $0.9$ & $3.5 \times 10^{5}$ & $5.4 \times 10^{4}$ & $5.5 \times 10^{4}$ & $7.9 \times 10^{4}$ & $2.2 \times 10^{-2}$ & $4.61 \times 10^{3}$ \\ \hline
$1.0 \times 10^{7}$ & $0.4$ & $0.8$ & $3.6 \times 10^{5}$ & $4.2 \times 10^{4}$ & $4.4 \times 10^{4}$ & $9.4 \times 10^{4}$ & $1.8 \times 10^{-1}$ & $1.68 \times 10^{3}$ \\ \hline
$1.0 \times 10^{7}$ & $0.4$ & $0.9$ & $3.5 \times 10^{5}$ & $4.2 \times 10^{4}$ & $4.4 \times 10^{4}$ & $9.2 \times 10^{4}$ & $1.5 \times 10^{-1}$ & $1.70 \times 10^{3}$ \\ \hline
\hline
$1.0 \times 10^{8}$ & $0.3$ & $0.8$ & $1.1 \times 10^{6}$ & $5.4 \times 10^{4}$ & $5.5 \times 10^{4}$ & $2.9 \times 10^{5}$ & $7.9 $ & $3.65 \times 10^{3}$ \\ \hline
$1.0 \times 10^{8}$ & $0.3$ & $0.9$ & $1.1 \times 10^{6}$ & $5.4 \times 10^{4}$ & $5.5 \times 10^{4}$ & $2.8 \times 10^{5}$ & $6.7 $ & $3.67 \times 10^{3}$ \\ \hline
$1.0 \times 10^{8}$ & $0.4$ & $0.8$ & $1.1 \times 10^{6}$ & $4.2 \times 10^{4}$ & $4.6 \times 10^{4}$ & $3.7 \times 10^{5}$ & $5.0 \times 10^{1}$ & $1.13 \times 10^{3}$ \\ \hline
$1.0 \times 10^{8}$ & $0.4$ & $0.9$ & $1.1 \times 10^{6}$ & $4.2 \times 10^{4}$ & $4.5 \times 10^{4}$ & $3.6 \times 10^{5}$ & $4.4 \times 10^{1}$ & $1.16 \times 10^{3}$ \\ \hline
\hline
$1.0 \times 10^{9}$ & $0.3$ & $0.8$ & $3.6 \times 10^{6}$ & $5.4 \times 10^{4}$ & $5.7 \times 10^{4}$ & $1.2 \times 10^{6}$ & $2.5 \times 10^{3}$ & $2.24 \times 10^{3}$ \\ \hline
$1.0 \times 10^{9}$ & $0.3$ & $0.9$ & $3.5 \times 10^{6}$ & $5.4 \times 10^{4}$ & $5.4 \times 10^{4}$ & $1.2 \times 10^{6}$ & $2.7 \times 10^{3}$ & $3.26 \times 10^{3}$ \\ \hline
$1.0 \times 10^{9}$ & $0.4$ & $0.8$ & $3.6 \times 10^{6}$ & $4.2 \times 10^{4}$ & $4.5 \times 10^{4}$ & $1.3 \times 10^{6}$ & $8.5 \times 10^{3}$ & $1.07 \times 10^{3}$ \\ \hline
$1.0 \times 10^{9}$ & $0.4$ & $0.9$ & $3.5 \times 10^{6}$ & $4.2 \times 10^{4}$ & $4.5 \times 10^{4}$ & $1.3 \times 10^{6}$ & $7.5 \times 10^{3}$ & $8.54 \times 10^{2}$ \\ \hline
\end{tabular}
\caption{Benchmark points for the $SU(2) \to U(1)$ symmetry breaking scenario for $n_f=0$, for a fixed mass parameter $m_0 = 10^4\,\mathrm{GeV}$. The table shows the evolution of transition strength parameter $\alpha_n$ with UV cutoff scale $m_X$ for various values of gauge coupling $g$ and scalar self-coupling $\lambda$.}
\label{tab:SU(2)PB_params}
\end{table}
Now we study the addition of fermions to the model. To this end, consider an $SU(2)$ gauge theory in which both the right-handed neutrino field $\Psi_R^a$ and a scalar field $\Phi^a$ transform in the adjoint (triplet) representation of $SU(2)$. One can write a gauge-invariant Yukawa term of the form
\begin{equation}
\mathcal{L}_\mathrm{Int.} \;=\;
 \frac{1}{2}\,y \,\epsilon^{abc} \,\bigl(\Psi_R^a\bigr)^T C \,\Psi_R^b \,\Phi^c
 \;+\;\text{h.c.},
\label{eq:YukawaTriplet}
\end{equation}
where $y$ is the Yukawa coupling constant, $C$ is the charge-conjugation matrix acting on the spinor indices, and $\epsilon^{abc}$ is the totally antisymmetric tensor contracting the $SU(2)$ indices $a,b,c$. Once the scalar field $\Phi^a$ acquires a VEV, the interaction in Eq.~(\ref{eq:YukawaTriplet}) induces a Majorana mass for the
right-handed neutrino. Plugging $\langle \Phi^3\rangle = v$ into
$\mathcal{L}_\mathrm{int}$ gives,
\begin{equation}
\frac{1}{2}\,y \,\epsilon^{ab3}\,(\Psi_R^a)^T C \,\Psi_R^b\,\langle \Phi^3\rangle
\;=\;
\frac{1}{2}\,y \,v \,\epsilon^{ab3}\,(\Psi_R^a)^T C \,\Psi_R^b,
\end{equation}
indicating that the resulting right-handed neutrino mass is proportional to the
product $m_{R_M} = y\,v$. This mechanism provides a direct connection between the scale of gauge symmetry breaking and neutrino mass generation, making it particularly relevant for seesaw models.

The possible representations for $N_R$ could be 3, 2, or 1. Choosing $N_R$ in the triplet (3) representation gives the following mass for neutrinos:
\begin{equation}
m_{R_M} = v.
\end{equation}
Once the $SU(2)$ group is broken, we have the same form of the potential as in Eq.~(\ref{eq:general_Lag}). Since $SU(2)$ is broken to $U(1)$, we have two massive gauge bosons, each with three degrees of freedom, hence 6 bosonic degrees of freedom. Considering three families of right-handed neutrinos, each in a triplet representation, we have $9 \times 2 = 18$ fermionic degrees of freedom.
\begin{table}[h!]
\centering
\renewcommand{\arraystretch}{1.5}
\begin{tabular}{|c|c|c|c|c|c|c|c|c|}
\toprule
\hline
\multicolumn{3}{|c|}{\textbf{Model parameters}} & \textbf{VEV} & \multicolumn{3}{c|}{\textbf{Temperatures}} & \multicolumn{2}{c|}{\textbf{Phase Trans.}} \\ \hline
$m_X$ & $g$ & $\lambda$ & $v$ & $T_0$ & $T_n$ & $T_c$ & $\alpha_n$ & $\beta/H_n$ \\ \hline
 [GeV] &  &  & [GeV] & [GeV] & [GeV] & [GeV] & &  \\ \hline \hline
$1.0 \times 10^{7}$ & $0.3$ & $0.8$ & $3.6 \times 10^{5}$ & $8.7 \times 10^{4}$ & $8.7 \times 10^{4}$ & $9.0 \times 10^{4}$ & $6.4 \times 10^{-4}$ & $9.34 \times 10^{4}$ \\ \hline
$1.0 \times 10^{7}$ & $0.3$ & $0.9$ & $3.5 \times 10^{5}$ & $8.7 \times 10^{4}$ & $8.7 \times 10^{4}$ & $8.9 \times 10^{4}$ & $5.8 \times 10^{-4}$ & $9.83 \times 10^{4}$ \\ \hline
$1.0 \times 10^{7}$ & $0.4$ & $0.8$ & $3.6 \times 10^{5}$ & $6.7 \times 10^{4}$ & $6.8 \times 10^{4}$ & $7.6 \times 10^{4}$ & $3.9 \times 10^{-3}$ & $2.60 \times 10^{4}$ \\ \hline
$1.0 \times 10^{7}$ & $0.4$ & $0.9$ & $3.5 \times 10^{5}$ & $6.7 \times 10^{4}$ & $6.8 \times 10^{4}$ & $7.6 \times 10^{4}$ & $3.5 \times 10^{-3}$ & $2.58 \times 10^{4}$ \\ \hline
\hline
$1.0 \times 10^{8}$ & $0.3$ & $0.8$ & $1.1 \times 10^{6}$ & $8.7 \times 10^{4}$ & $8.7 \times 10^{4}$ & $1.4 \times 10^{5}$ & $5.9 \times 10^{-2}$ & $9.61 \times 10^{4}$ \\ \hline
$1.0 \times 10^{8}$ & $0.3$ & $0.9$ & $1.1 \times 10^{6}$ & $8.7 \times 10^{4}$ & $8.7 \times 10^{4}$ & $1.4 \times 10^{5}$ & $5.1 \times 10^{-2}$ & $9.22 \times 10^{4}$ \\ \hline
$1.0 \times 10^{8}$ & $0.4$ & $0.8$ & $1.1 \times 10^{6}$ & $6.7 \times 10^{4}$ & $6.7 \times 10^{4}$ & $1.9 \times 10^{5}$ & $7.5 \times 10^{-1}$ & $2.98 \times 10^{4}$ \\ \hline
$1.0 \times 10^{8}$ & $0.4$ & $0.9$ & $1.1 \times 10^{6}$ & $6.7 \times 10^{4}$ & $6.7 \times 10^{4}$ & $1.8 \times 10^{5}$ & $6.4 \times 10^{-1}$ & $3.08 \times 10^{4}$ \\ \hline
\hline
$1.0 \times 10^{9}$ & $0.3$ & $0.8$ & $3.6 \times 10^{6}$ & $8.7 \times 10^{4}$ & $8.7 \times 10^{4}$ & $6.3 \times 10^{5}$ & $4.1 \times 10^{1}$ & $3.35 \times 10^{4}$ \\ \hline
$1.0 \times 10^{9}$ & $0.3$ & $0.9$ & $3.5 \times 10^{6}$ & $8.7 \times 10^{4}$ & $8.7 \times 10^{4}$ & $6.0 \times 10^{5}$ & $3.4 \times 10^{1}$ & $3.43 \times 10^{4}$ \\ \hline
$1.0 \times 10^{9}$ & $0.4$ & $0.8$ & $3.6 \times 10^{6}$ & $6.7 \times 10^{4}$ & $6.7 \times 10^{4}$ & $9.4 \times 10^{5}$ & $6.1 \times 10^{2}$ & $1.17 \times 10^{4}$ \\ \hline
$1.0 \times 10^{9}$ & $0.4$ & $0.9$ & $3.5 \times 10^{6}$ & $6.7 \times 10^{4}$ & $6.7 \times 10^{4}$ & $9.0 \times 10^{5}$ & $5.1 \times 10^{2}$ & $1.19 \times 10^{4}$ \\ \hline
\end{tabular}
\caption{Benchmark points for the $SU(2) \to \mathds{1}$ symmetry breaking scenario without fermions ($n_f=0$) for a fixed mass parameter $m_0 = 10^4\,\mathrm{GeV}$. Complete breaking of $SU(2)$ produces three massive gauge bosons, leading to stronger thermal effects compared to partial breaking.}
\label{tab:SU(2)MB_params}
\end{table}
In the left panel of \fig{fig:SU(2)_all} we plot the thermally corrected potential, whereas in the case of $U(1)$, \fig{fig:U(1)_potentials} we can identify the thermal origin of the barrier for different scales, $m_X=10^7, 10^8, 10^9$ GeV. As expected, the more degrees of freedom, the higher the potential barrier, with the partial breaking $SU(2) \to U(1)$ producing roughly three times stronger thermal effects compared to the $U(1)$ case for the same choice of couplings and scales (compare Figs.~(\ref{fig:U(1)_potentials}) and (\ref{fig:SU(2)_all})). This enhanced barrier strength arises from the larger number of gauge degrees of freedom contributing to thermal loops. Then, in \tab{tab:SU(2)PB_params} we present some benchmark points for the case of no fermions.  For the thermally corrected potential, we use again \eq{eq:Veff} with \eq{eq:VT}, using the complete numerical form of the thermal functions $J_B$ and $J_F$.  In Section \ref{sec:ResFOPT} we present the results when also fermions are included. 
\subsection{$SU(2) \longrightarrow \mathds{1}$}
While the $SU(2)\to U(1)$ case preserves a residual gauge symmetry, the complete breaking $SU(2)\to \mathds{1}$ represents the maximal scenario in which all gauge bosons acquire mass. 
This leads to the largest number of massive degrees of freedom (three gauge bosons with three polarizations each), further enhancing the thermal contributions that drive barrier formation. 
At the same time, the representation structure of the scalar field is different: the breaking is achieved through the fundamental representation rather than the adjoint. 
These features make the $SU(2)\to \mathds{1}$ case both phenomenologically richer and technically distinct, and we now present its formal construction.

In this case, the Lagrangian is given by:
\begin{align}
    \mathcal{L}=(D^\mu \Phi)^\dagger (D_\mu \Phi) -V(\Phi)-\frac{1}{4} X_{\mu \nu}^a X^{a \mu \nu}.
\end{align}
Here, the first term represents the kinetic energy and gauge interactions of the scalar field, the second term is the scalar potential, and the third term describes the dynamics of the gauge fields. $\Phi$ is now an $SU(2)$ scalar doublet, and $X^a_{\mu\nu}$ is the field strength tensor for the $SU(2)$ gauge fields $A_\mu^a$. The field strength tensor is defined as follows:
\begin{align}
    X^a_{\mu\nu} = \partial_\mu A^a_\nu - \partial_\nu A^a_\mu + g \epsilon^{abc} A^b_\mu A^c_\nu,
\end{align}
where $g$ is the gauge coupling constant and $\epsilon^{abc}$ is the Levi-Civita symbol. The covariant derivative acting on the scalar field is given by the usual expression:
\begin{align}
    D_\mu \Phi=\left(\partial_\mu + i g A_\mu^a \tau^a\right) \Phi,
\end{align}
where $\tau^a$ are the generators of $SU(2)$ defined in \eq{eq:paulimatrices} with $a=1,2,3$. 
Since $\Phi$ is composed of complex scalar fields, it contains four real degrees of freedom. In the unitary gauge, and aligning the vev to the down component, the scalar field takes the form
\begin{align}
    \Phi(x) = \frac{1}{\sqrt{2}} \begin{pmatrix}
        0 \\
        v+h(x)
    \end{pmatrix},
\end{align}
where $v$ is the vacuum expectation value (VEV) of the scalar field and $h(x)$ is the fluctuation of the field. The non-trivial minimum of the potential $V_\text{Tree}(\Phi)$ occurs when the scalar field acquires a VEV 
\begin{align}
\label{eq:usualphivev}
    \langle \Phi \rangle = \frac{1}{\sqrt{2}} \begin{pmatrix}
        0 \\ 
        v
    \end{pmatrix}.
\end{align}
To find the masses of the gauge bosons, we follow the standard procedure of substituting the vev of \ref{eq:usualphivev} into the kinetic term of the Lagrangian. We find that the gauge bosons acquire mass term
\begin{align}
    \left(D_\mu \langle\Phi\rangle \right)^{\dagger} D^\mu \langle\Phi\rangle = \frac{g^2v^2}{8} A_\mu^a A^{a\mu} = \frac{m_b^2}{2} A_\mu^a A^{a\mu},
\end{align}
where the mass of the gauge bosons is then given by:
\begin{align}
    m_b = \frac{g}{2}v.
\end{align}
The three gauge bosons $A_\mu^1, A_\mu^2$ and $A_\mu^3$ get the same mass due to the complete spontaneous breaking of the gauged $SU(2)$ symmetry. This symmetry breaking generates three Nambu-Goldstone bosons, corresponding to the three broken generators of $SU(2)$. These Goldstone bosons are eaten by the gauge fields to provide the longitudinal polarization states necessary for massive vector bosons. As a result, each massive gauge boson has three degrees of freedom, giving a total of nine degrees of freedom for the three massive bosons.

Now, let us consider the case with fermions. The fundamental representation of $SU(2)$ severely constrains the possible fermion interactions due to the structure of the gauge group. If both left-handed and right-handed fermion are $SU(2)$ doublets, we cannot generate mass through the Yukawa term.
 This is because the Yukawa term between the $SU(2)$ doublet scalar field and $SU(2)$ doublet fermions cannot form a gauge singlet. 

 More generally, for Majorana fermions, we need to find a representation $R$ such that the tensor product $2 \otimes R \otimes R$ contains a singlet, where $2$ refers to the fundamental doublet representation of the scalar field. Hence, in general, if we try to add Majorana fermions, we would need to find a representation $R$ such that the product 
\bea
2 \otimes  R \otimes R,
\eea
contains a singlet. This, however, is not possible. This fundamental obstruction can be understood from the representation theory of $SU(2)$. To see it, we can calculate the tensor product with $V_j\bigotimes V_m = V_{j+m} \bigoplus V_{j+m-1}\bigoplus \hdots \bigoplus V_{j-m+1}\bigoplus V_{j-m}$, here the dimension of the representation is $d=2j+1$, for $j=0,1/2, 3/2,\hdots$, for example
\bea
2 \otimes 2 = 3 \oplus 1, \nonumber\\
2 \otimes 3 = 4 \oplus 2, \\
3 \otimes 3 = 5 \oplus 3  \oplus 1, \nonumber\\
4 \otimes 4  = 7 \oplus 5 \oplus 3 \oplus 1.
\eea
Hence, using the familiar notation of spins, for the triple product to contain a singlet ($J_{\text{tot}}=0$), two representations must be able to couple into some intermediate spin $J$ such that
$\tfrac{1}{2} \otimes J \;\supset\; 0.$, but this is only possible if $J=\tfrac{1}{2}$.
The decomposition of $R_2 \otimes R_2$, where $R_2$ corresponds to spin $j_2$: 
$j_2 \otimes j_2 \;=\; \bigoplus_{J=0}^{2j_2} J$. Note that this direct sum contains only integer spins $J$, since $2j_2$ is always an integer. Therefore, the representation $J=\tfrac{1}{2}$ (corresponding to 2) never appears in $j_2 \otimes j_2$. Consequently, when coupling with the doublet $j=\tfrac{1}{2}$, one can never obtain a singlet:
\bea
\tfrac{1}{2} \otimes (j_2 \otimes j_2) \;\not\supset\; 0.
\eea
This mathematical constraint means that, unlike the $SU(2) \to U(1)$ case, we cannot easily incorporate Majorana neutrino masses in the complete breaking scenario, which limits its direct connection to neutrino mass generation via the seesaw mechanism.

However, we can still consider the case where right-handed fermions are $SU(2)$ singlets, as in the Standard Model. In this case, the Yukawa term becomes gauge invariant because the product of a left-handed doublet, the scalar doublet, and a right-handed singlet can form a gauge singlet:
\begin{align}
    \mathcal{L}_\text{Yukawa} &= - \frac{y}{\sqrt{2}} \left(\bar{\Psi}_L \Phi \psi_{R,2} + \bar{\psi}_{R,2} \Phi^\dagger \Psi_L + \bar{\Psi}_L \tilde{\Phi} \psi_{R,1} + \bar{\psi}_{R,1} \tilde{\Phi}^\dagger \Psi_L \right) 
\end{align}
where $\tilde{\Phi} = i\sigma^2\Phi^*$ is the charge-conjugate of the scalar doublet. The fermion masses generated from this Yukawa term at VEV are as follows:
\begin{align}
    m_f = \frac{y}{\sqrt{2}}v.
\end{align}
In the right panel of \fig{fig:SU(2)_all} we present the shape of thermally corrected potential, where again we can see the thermal origin of the barrier for different scales, $m_X=10^7, 10^8, 10^9$ GeV,  while keeping $m_0=10^4$ GeV. Then, in \tab{tab:SU(2)MB_params} we present some benchmark points.  For the thermally corrected potential we have used \eq{eq:Veff} with \eq{eq:VT}, using the complete numerical form of the thermal functions $J_B$ and $J_F$, \eq{eq:JBJF}, and the thermal energy corrections given in Appendix \ref{app:furthercontThPot}. 

The parameter values in our benchmark tables span several orders of magnitude in the transition strength $\alpha_n$ (from $10^{-3}$ to $10^3$), reflecting the sensitivity of the phase transition to the UV scale $m_X$, and the coupling strengths. The inverse duration parameter $\beta/H_n$ shows corresponding variations, with stronger transitions (larger $\alpha_n$) generally proceeding faster. These wide parameter ranges demonstrate the rich phenomenology accessible in flat potential scenarios and highlight the importance of thermal effects in determining gravitational wave signatures.

\subsection{Scaling Behavior \label{subsec:scaling}}

To understand why the phase transition becomes stronger at higher UV scales $m_X$, we now analyze the scaling of thermal, loop, and tree-level contributions.

The emergence and growth of the barrier in the effective potential can be traced to a clear hierarchy between thermal/loop-induced effects and the tree-level sextic interaction. At finite temperature, integrating out fluctuations around the scalar field generates an effective quartic term, which scales as
\begin{equation}
\lambda_4^{\rm (thermal)} \sim \lambda_6 \frac{T^2}{m_X^2}.
\end{equation}
This contribution arises from thermal loops involving the dimension-six operator and reflects the enhancement of scalar fluctuations in the plasma. However, since it is inversely proportional to \(m_X^2\), this term becomes increasingly negligible for large values of \(m_X\). A similar conclusion follows from the quantum loop correction to the quartic, which scales as
\begin{equation}
\Delta \lambda_4^{\rm (loop)} \sim \frac{\lambda_6 m_0^2}{m_X^2} \frac{1}{16\pi^2} \ln \frac{m_X^2}{m_0^2}.
\end{equation}
Although the logarithm grows slowly with $m_X$, the prefactor $m_0^2/m_X^2$ ensures that this correction also diminishes in the large-$m_X$ limit. In contrast, the dominant origin of the barrier is the classical sextic interaction present at tree level:
\begin{equation}
V(\phi) \supset \frac{\lambda_6}{720 m_X^2} \phi^6.
\end{equation}
Despite being suppressed by $1/m_X^2$, this term becomes increasingly important at large field values. When combined with the quadratic mass term, the resulting potential develops a nontrivial minimum away from the origin, located approximately at
\begin{equation}
\phi_{\min}^4 \sim m_0^2 m_X^2,
\end{equation}
for temperatures near \(T \sim m_0\). This field value grows with \(m_X\), and the barrier height associated with this secondary minimum scales as
\begin{equation}
\label{eq:VBarrier}
V_{\rm Barrier} \sim \alpha \phi_{\min}^6 \sim m_0^3 m_X.
\end{equation}
Thus, the barrier becomes significantly taller as $m_X$ increases, despite the sextic term's inverse dependence on $m_X^2$. This apparent paradox is resolved by noting that the field displacement $\phi_{\min}$ grows with $m_X$, effectively enhancing the impact of the sextic operator at large field values. Moreover, expanding the potential around the nontrivial vacuum generates an effective quartic coupling given by
\begin{equation}
\lambda_{4,\rm Eff. }^{\rm (Tree)} \sim \frac{\lambda_6}{2} \frac{v^2}{m_X^2},
\end{equation}
where $v = \phi_{\min} \sim \sqrt{m_0 m_X}$, leading to $\lambda_{4,\rm Eff. }^{\rm (Tree)} \sim \lambda_6 \, m_0/m_X$. Importantly, this result is approximately independent of $m_X$ , confirming that the increasing strength of the phase transition is not due to a strengthening of the effective quartic but rather due to the increasing height and width of the barrier, both of which scale with $m_X$.

In summary, while thermal and quantum corrections generate quartic terms that diminish with increasing $m_X$ , the classical structure of the potential, especially the tree-level sextic operator, drives the formation and growth of the barrier. The key effect is that the minimum shifts to larger field values as $m_X$  increases, amplifying the contribution of the sextic term and resulting in a barrier whose height scales as $V_{\rm Barrier} \sim m_0^3 m_X$ , \eq{eq:VBarrier}. This confirms that the strong first-order character of the phase transition at large $m_X$ is fundamentally a tree-level phenomenon.

\section{Results on First Order Phase Transitions and Gravitational Waves \label{sec:ResFOPT}}
FOPTs emerge from the interplay between thermal fluctuations restoring symmetry and quantum corrections favoring the broken phase, with effective flat potentials leading to gravitational wave signatures. The resulting spectrum is determined by nucleation, expansion, collisions, and plasma response, and is primarily controlled by the transition strength $\alpha$, the inverse duration $\beta/H$, the nucleation temperature $T_n$ and the bubble wall velocity . The dynamics are described through the effective potential \eq{eq:Veff}, with three main regimes: non-runaway bubbles, non-runaway bubbles in plasma, and vacuum transitions with $\alpha>1$ \footnote{For an up-to-date review of FOPT we consulted the latest LISA Cosmology Working Group document~\cite{Caprini:2024hue}. For general reviews on FOPT, see, for example, \cite{Hindmarsh:2021l, Caprini:2020LI,Athron:2023xlk}}. Each case exhibits different hydrodynamical behavior and relative contributions from sound waves, bubble collisions, scalar fields, and MHD turbulence, shaping the GW spectrum. We use the most up-to-date fits available, noting that further developments in simulations may reveal additional features not yet known to the community.
\begin{figure}[h]
    \centering
    \begin{subfigure}[b]{0.49\textwidth}
        \includegraphics[width=\textwidth]{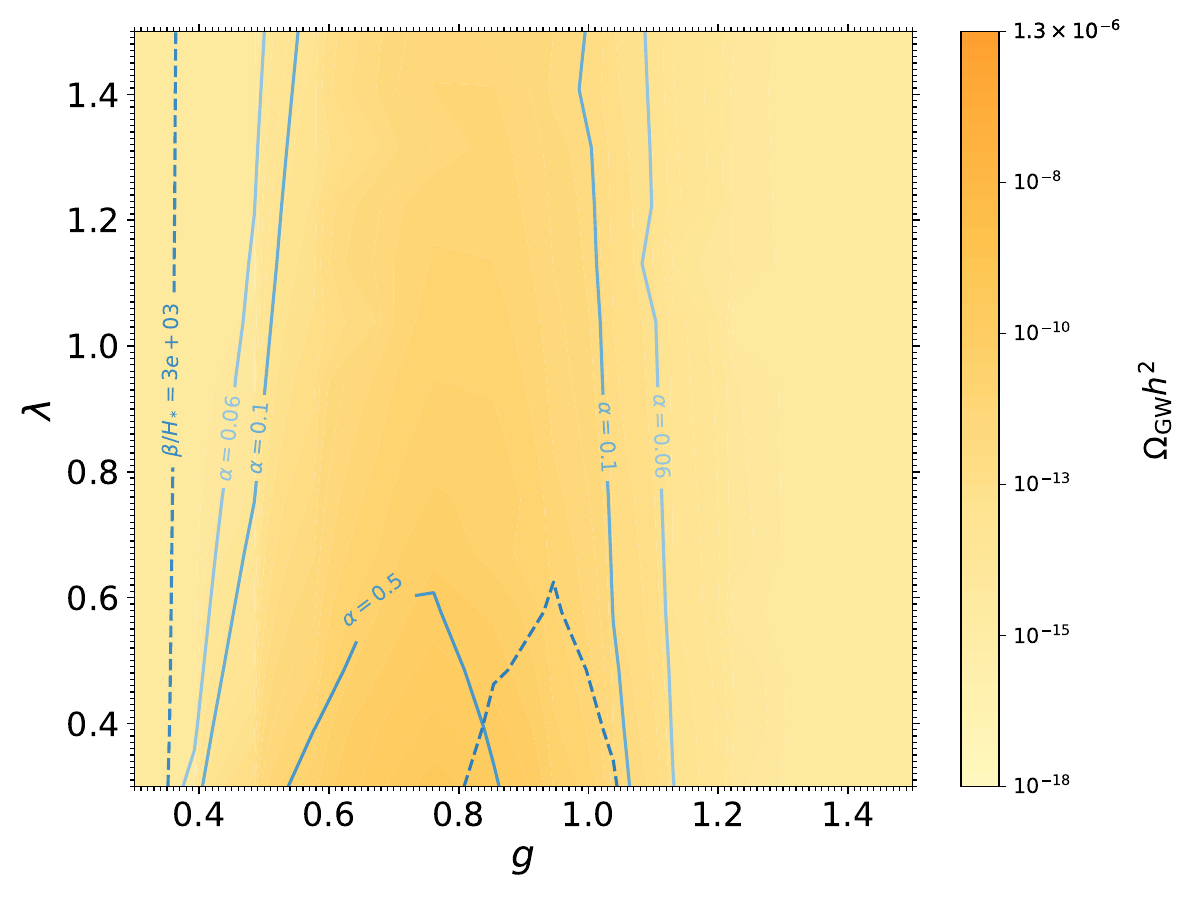}
        \label{fig:grid_U1_mX1e7}
    \end{subfigure}
    \hfill
    \begin{subfigure}[b]{0.49\textwidth}
        \includegraphics[width=\textwidth]{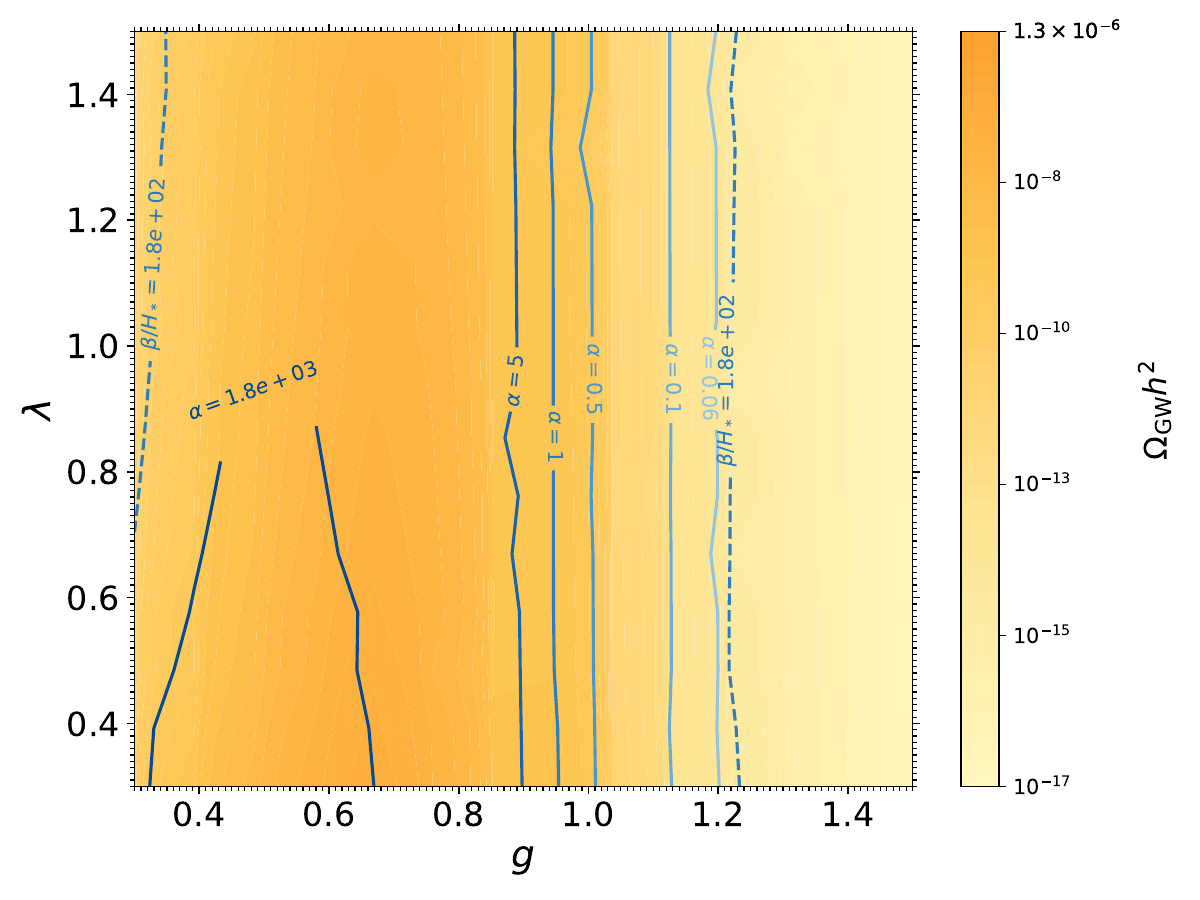}
        \label{fig:grid_U1_mX1e9}
    \end{subfigure}
    \caption{Phase transition parameters for $U(1)$ symmetry breaking in the $(g, \lambda)$ parameter space and the corresponding GW density. \textbf{Left:} At $m_X = 10^7$ GeV, and $n_f=0$, showing contours for $\alpha$ (solid lines) and $\beta/H_*$ (dashed lines). \textbf{Right:} At $m_X = 10^9$ GeV, contours for $\alpha$ and $\beta/H_*$. The color gradient indicates the corresponding GW density.}
    \label{fig:grid_GWU1}
\end{figure}
\begin{figure}
    \centering
 \begin{subfigure}[b]{0.49\textwidth}
        \includegraphics[width=\linewidth]{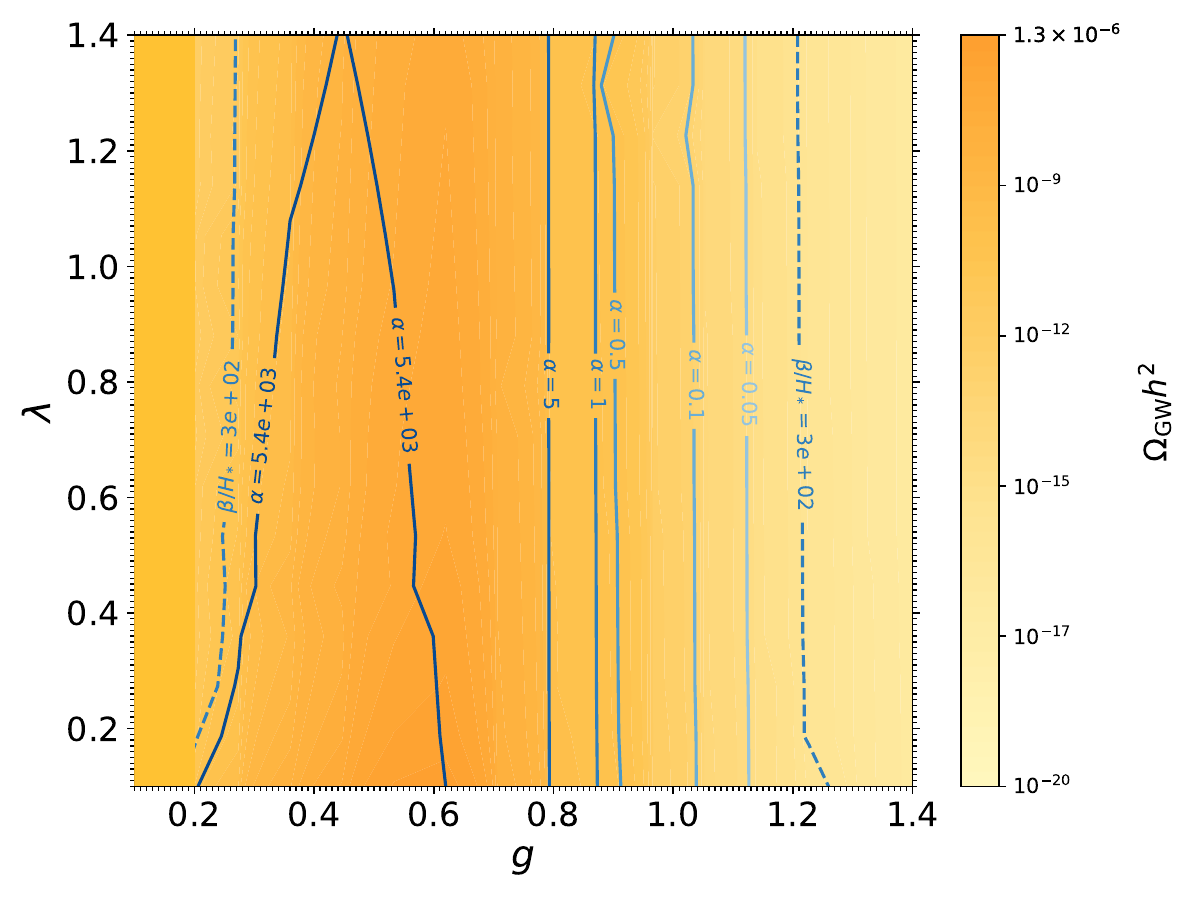}
        \caption{$n_f=0$, $y = 0.00$}
        \label{fig:scan1SU2U1}
    \end{subfigure}
    \hfill
    \begin{subfigure}[b]{0.49\textwidth}
        \includegraphics[width=\linewidth]{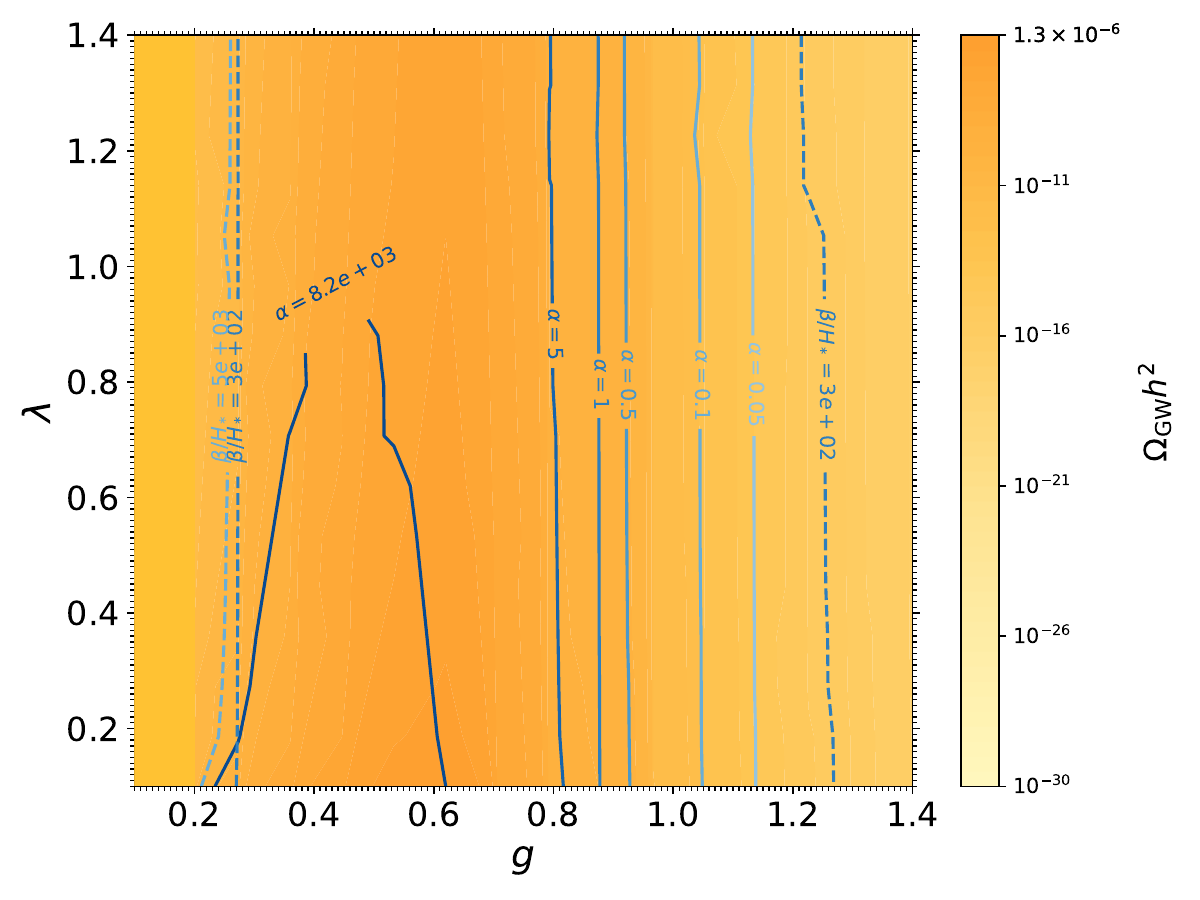}
        \caption{$n_f = 6$, $y = 0.35$}
        \label{fig:scanSU22U1}
    \end{subfigure}
    \caption{Comparison of scan results for different $n_f=0$ (left) and $n_f = 6$, $y = 0.35$ (right) values, for the case of the breaking $SU(2)\rightarrow U(1)$.}
    \label{fig:scansSU2U1_A}
\end{figure}
\begin{figure}
    \centering
    \begin{subfigure}[b]{0.49\textwidth}
        \includegraphics[width=\linewidth]{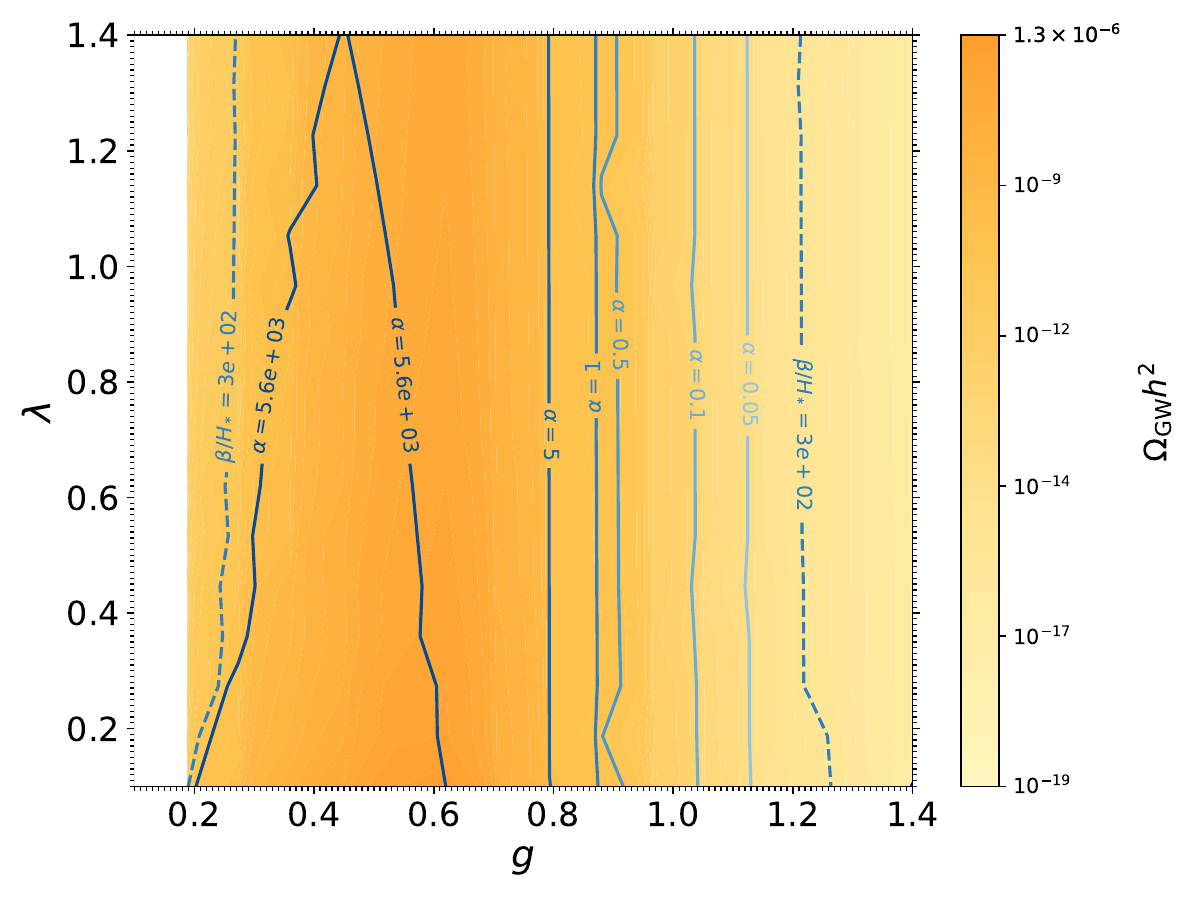}
        \caption{$n_f=12$, $y = 0.1$}
        \label{fig:scanSU2U1_3}
    \end{subfigure}
    \hfill
    \begin{subfigure}[b]{0.49\textwidth}
        \includegraphics[width=\linewidth]{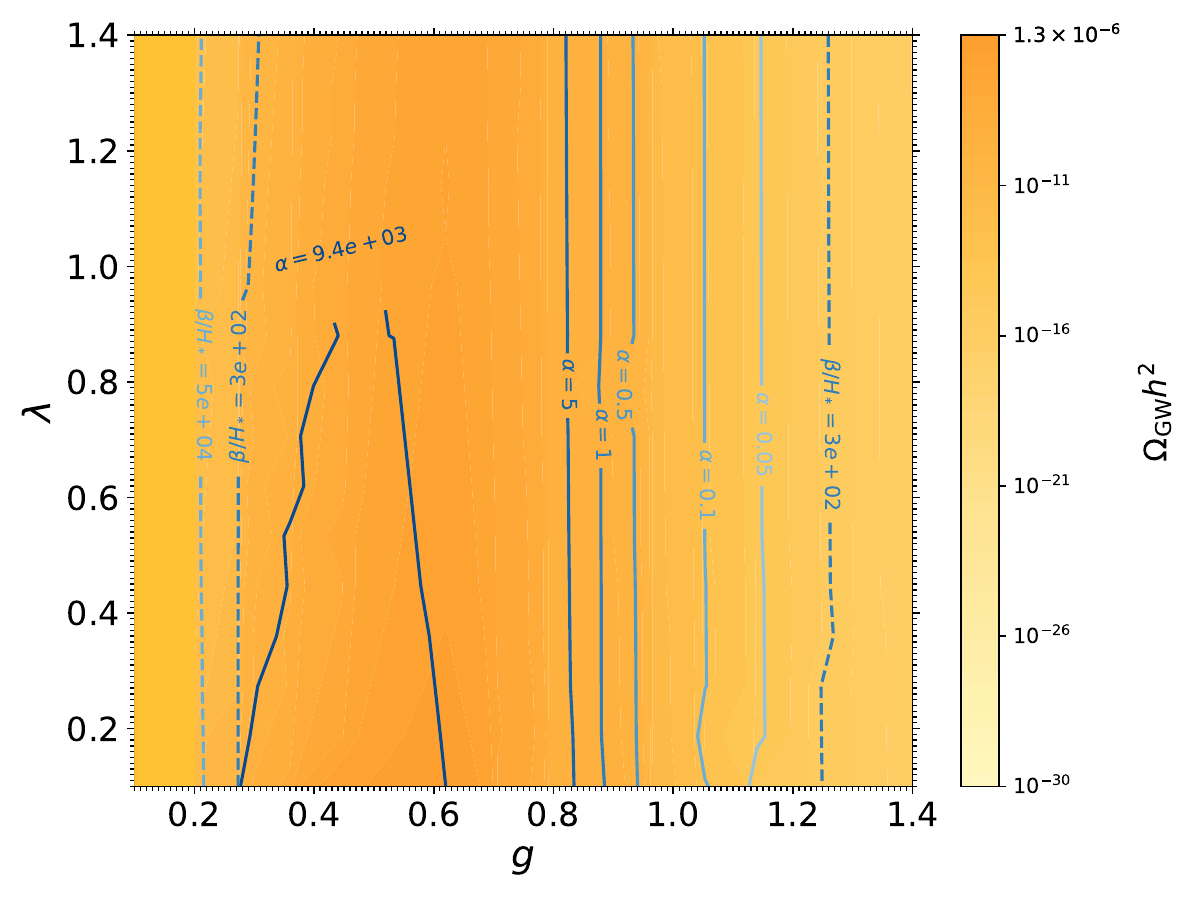}
        \caption{$n_f = 12$, $y = 0.35$}
        \label{fig:scanSU2U1_4}
    \end{subfigure}

    \vskip\baselineskip

    \begin{subfigure}[b]{0.48\textwidth}
        \includegraphics[width=\linewidth]{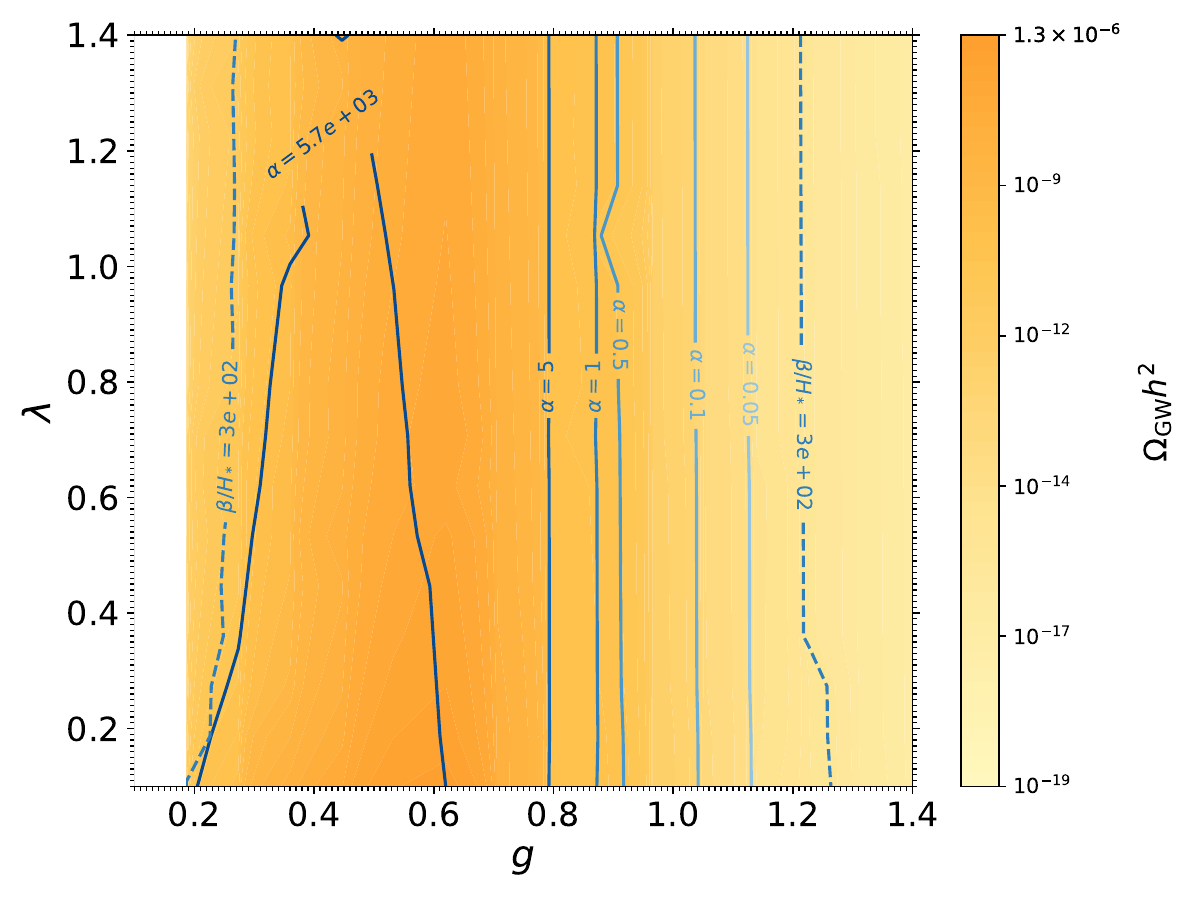}
        \caption{$n_f = 18$, $y = 0.10$}
        \label{fig:scanSU2U1_5}
    \end{subfigure}
    \hfill
    \begin{subfigure}[b]{0.48\textwidth}
        \includegraphics[width=\linewidth]{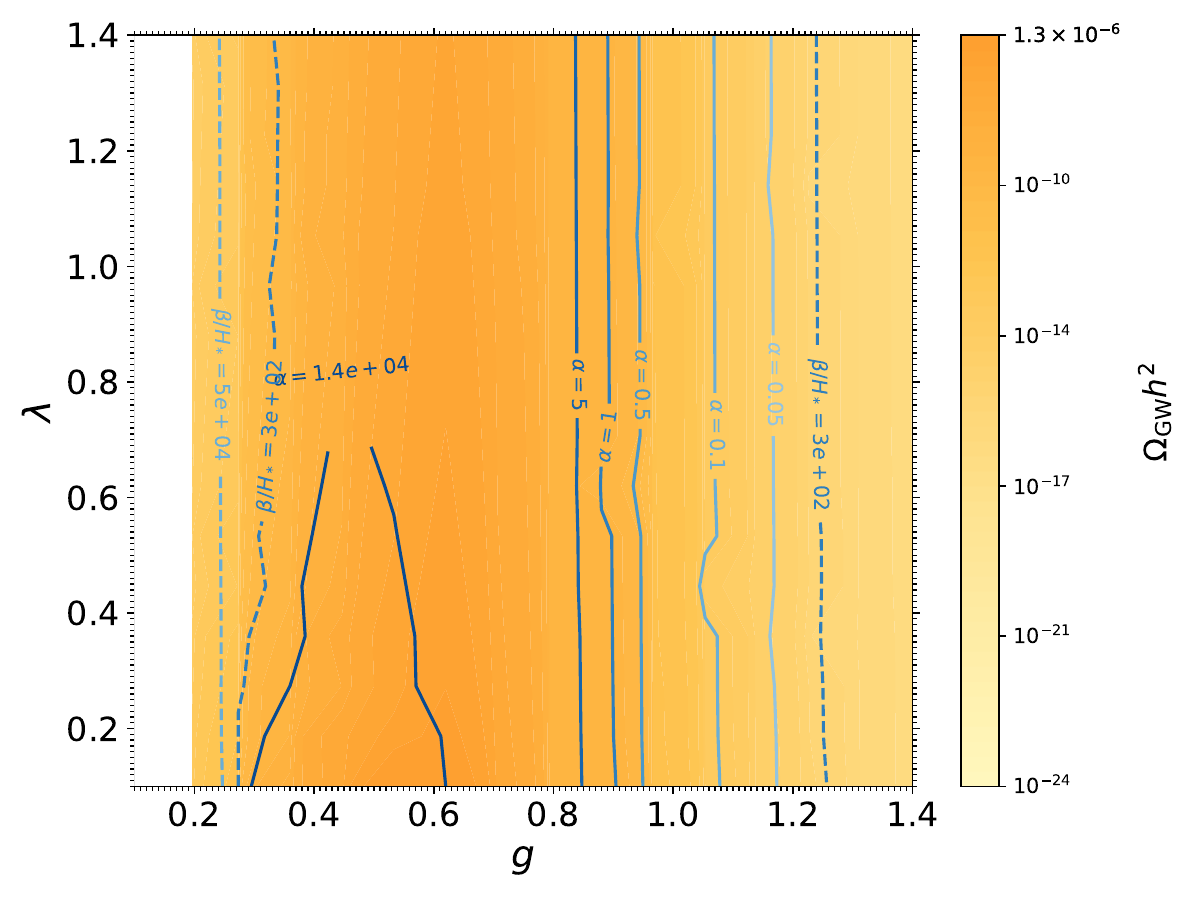}
        \caption{$n_f = 18$, $y = 0.35$}
        \label{fig:scanSU2U1_6}
    \end{subfigure}

    \caption{Comparison of scan results for different $n_f$ and $y$ values.}
    \label{fig:scansSU2U1_B}
\end{figure}
\begin{figure}
    \centering
    \includegraphics[width=0.5\linewidth] {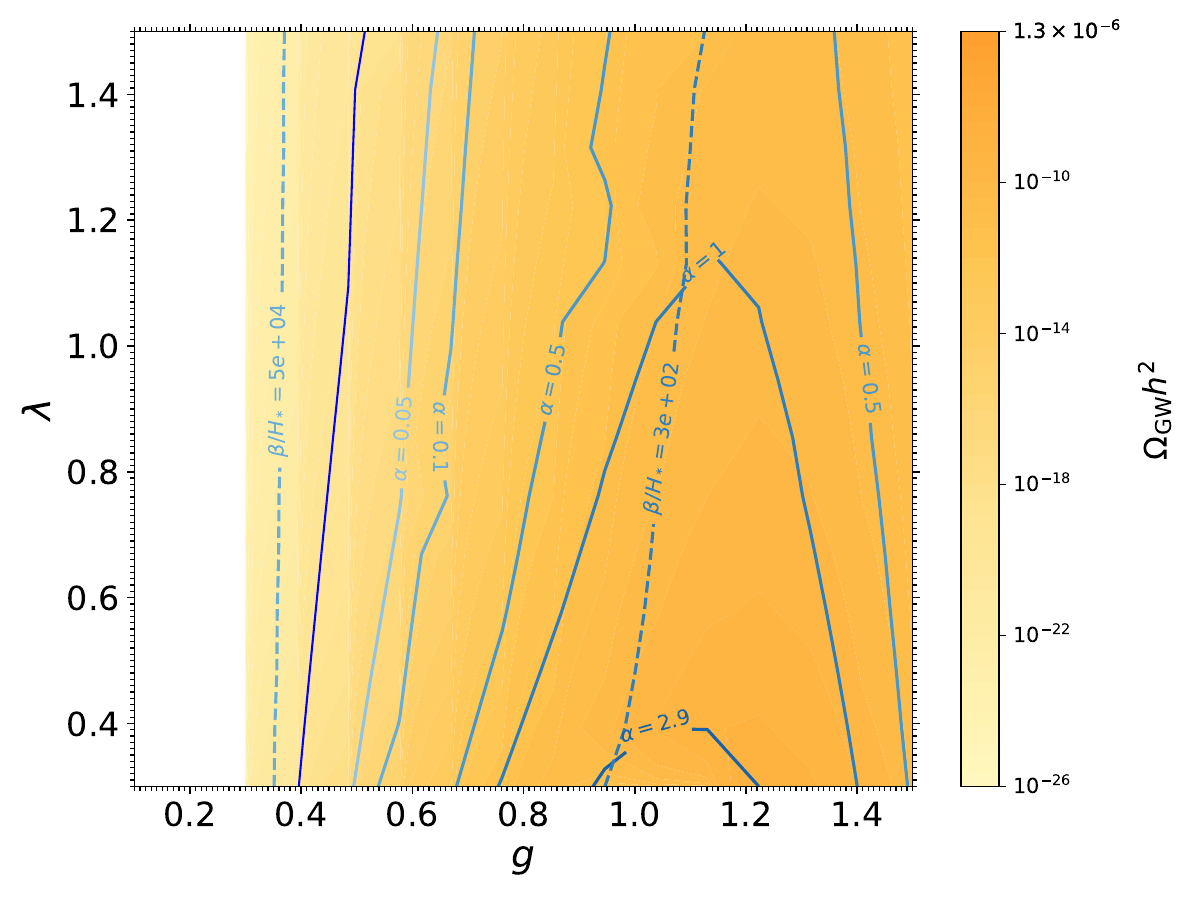}
    \caption{Phase transition parameters for $SU(2) \to \mathds{1}$ symmetry breaking in the ($g, \lambda$) space at $m_X = 10^7 \, \text{GeV}$ with the same format as Figure \ref{fig:grid_GWU1}. }
    \label{fig:scansSU2not}
\end{figure}
\subsection{Transition Strength and Duration Parameters}

The transition strength parameter $\alpha$ quantifies the ratio of vacuum energy released during the phase transition to the background radiation energy density:
\bea
\label{eq:alphadef}
\alpha=\frac{\text{latent\ heat}}{\rho_R}.
\eea
This parameter controls the violence of bubble expansion and collision dynamics. When $\alpha \ll 1$, bubbles expand sub-sonically and the transition proceeds gently, while $\alpha \sim 1$ or larger indicates relativistic bubble walls and violent collisions that efficiently convert vacuum energy into gravitational waves. The inverse duration parameter $\beta$ measures how rapidly the nucleation rate changes with temperature:
\bea
\label{eq:betadef}
\beta=H_n\left.T\frac{d}{dT}\left(\frac{S_3}{T} \right)\right|_{T= T_n}.
\eea
The ratio $\beta/H$ determines the typical timescale of the transition relative to the Hubble time. The lower values of $\beta/H$ correspond to longer transitions that allow more time for gravitational wave production, while the higher values indicate rapid completion with less efficient energy conversion.

Our flat potential scenarios naturally produce a wide range of these parameters, depending on the UV cut-off scale $m_X$ and coupling strengths. As shown in Tables~\ref{tab:U(1)_params}-\ref{tab:SU(2)MB_params}, increasing $m_X$ from $10^7$ to $10^9$ GeV dramatically enhances $\alpha$ from $\mathcal{O}(10^{-3})$ to $\mathcal{O}(10^3)$, reflecting the stronger potential barriers that develop at higher scales.
\subsection{Case of Radiation Domination}

Depending on the radiation domination, $\alpha<1$, or the vacuum domination time, $\alpha>1$, the expression for \eq{eq:alphadef} can be written in different ways. For the well known of radiation domination we have 
\bea
\alpha=\frac{1}{\rho_{\rm rad}}\left.\left[\Delta V(h_c,T)-\frac{T}{4}\frac{d\Delta V(h_c,T)}{dT}\right]\right|_{T=T_n} ,
\label{eq:alpha_radiation}
\eea
where the $\Delta V= V(\phi=0)-V(\phi=\phi_c)$, 
and the beta parameter is
 \bea
\frac{\beta}{H_*}=\left.T_n\frac{d}{dT}\left(\frac{S_3}{T} \right)\right|_{T=T_n\approx T_*},
\eea
since $T_n \approx T_R \approx T_*$ and so $H_*\approx H(T_n) \approx H(T_R)$.
In this regime, the universe remains radiation-dominated throughout the transition, and the standard relations between nucleation, completion, and reheating temperatures hold. This is the most commonly studied case and applies to transitions with moderate strength parameters where the vacuum energy release does not significantly alter the expansion history.

\subsection{Case of Vacuum Domination \label{sec:vac_dom}}
When $\alpha > 1$, the energy density locked in the false vacuum can temporarily dominate over radiation, leading to a brief period of vacuum-driven expansion. This scenario requires more careful treatment of the temperature evolution and timing relationships. 
A conservative bound on the density of GW produced is to limit the quantity via the number of effective neutrinos, using
\bea 
\Delta N_{\rm Eff. }^{\rm GW} \simeq \frac{8}{7}\left(\frac{11}{4}\right)^{4/3}\frac{\rho_{\rm GW}}{\rho_\gamma}
= \frac{8}{7}\left(\frac{11}{4}\right)^{4/3}\frac{\Omega_{\rm GW}^{\rm Tot}}{\Omega_\gamma}\,,
\eea 
where $\rho_{\rm GW}$ and $\rho_\gamma$ are respectively the GW and radiation densities.
For $\Delta N_{\rm Eff. }\lesssim \mathcal{O}(0.2)$ \cite{Planck2018}, we have for the total density
$\Omega_{\rm GW}^{\rm Tot} h^2 \lesssim \mathrm{few}\times 10^{-6}$, which is the bound that we consider.

\begin{figure}
    \centering
    \includegraphics[width=0.95\linewidth]{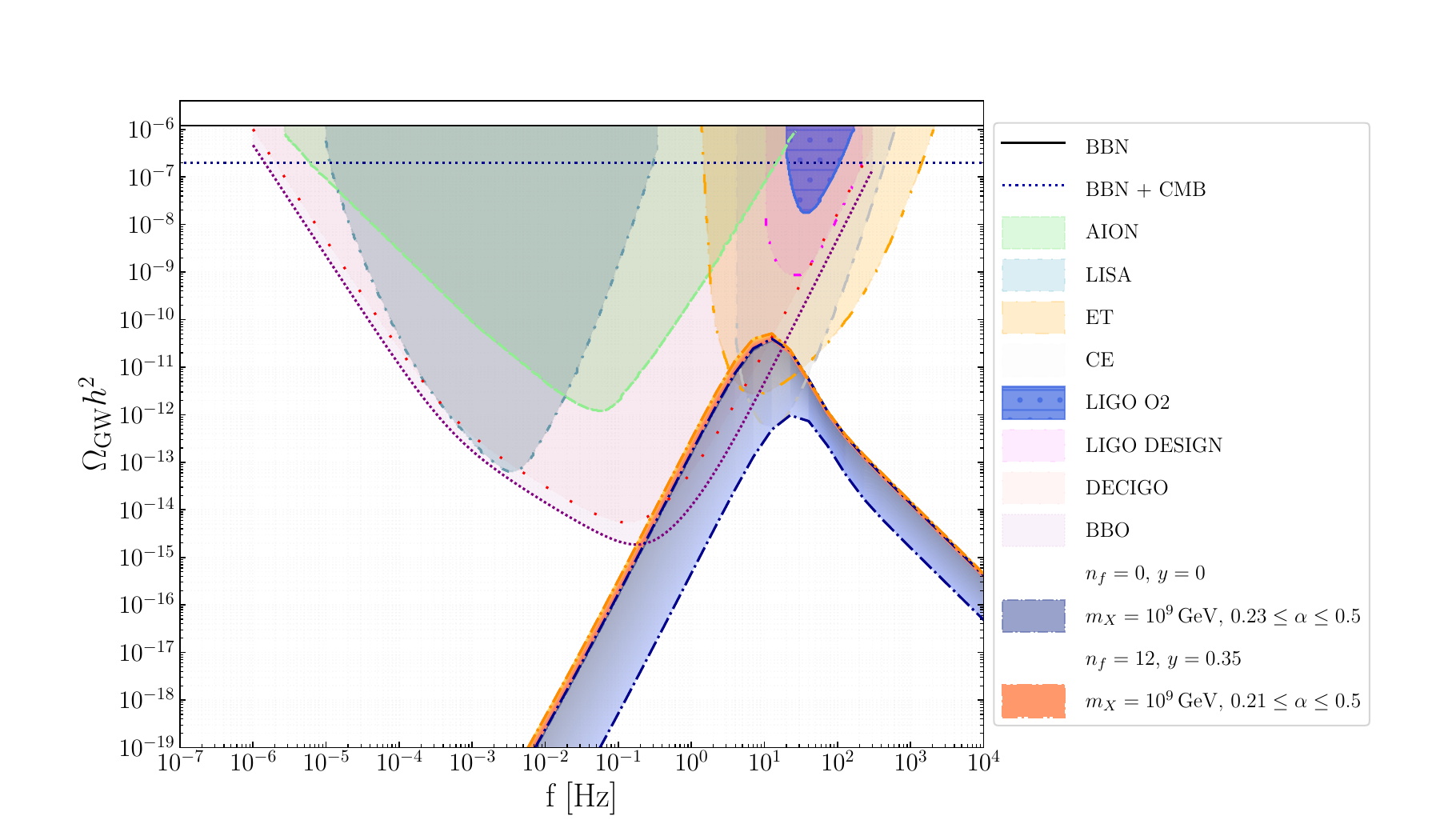}
    \caption{FOPT signatures of the $SU(2)\rightarrow U(1)$ model. For $m_X=10^9$ GeV, $n_f=0$, $y=0$, we plot GW densities for $\alpha=0.23$ up to $\alpha=0.5$ (blue shaded region), which correspond to reliable simulations. For $m_X=10^9$ GeV, $n_f=12$, $y=0.35$, we plot the density profiles for $\alpha=0.21 \leq \alpha \leq 0.5$, which overlap with DECIGO, BBO, ET. }
    \label{fig:GW_densitySU2U1}
\end{figure}

\begin{figure}
    \centering
    \includegraphics[width=0.95\linewidth]{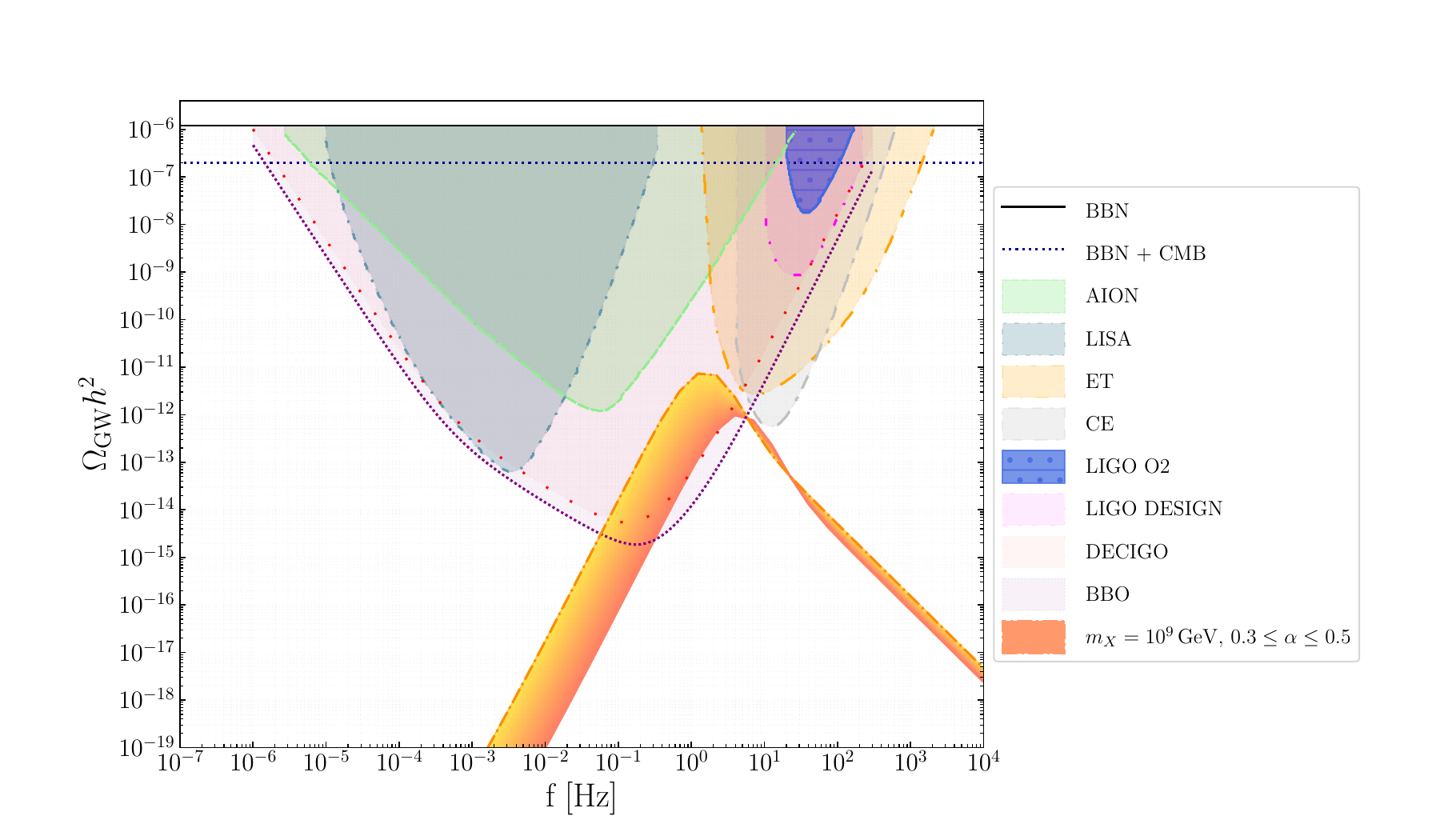}
    \caption{FOPT signatures of the $SU(2)\rightarrow \mathds{1}$ model that lie in an putative observable regions. The shaded graded orange area corresponds to $0.3 \leq \alpha\leq 0.5$.}
    \label{fig:GW_densitySU2not}
\end{figure}

The important point for the vacuum domination, which is characterized by runaway bubbles (in the regime of strongly supercooled FOPT \cite{Ellis:2018mja}) is that in general
\bea
\label{eq:temphierarchies}
T_n \ll T_R \approx T_*,
\eea
where $T_n$, $T_R$ and $T_*$ are respectively the nucleation temperature, the reheating temperature and the thermal bath temperature. Hence, in this case
\bea
\alpha=\frac{\rho_{\rm{vac.}}}{\rho_{\rm{rad.}}(T_n)}.
\eea
For $\beta$,  \eq{eq:betadef}, for this case when calculating the GW spectra we need to get it at  the end of production, which corresponds to $H_*$ (the Hubble size at the thermal bath temperature), hence
 \bea
\frac{\beta}{H_*}=\frac{H_n}{H_*}\left.T_n\frac{d}{dT}\left(\frac{S_3}{T} \right)\right|_{T=T_n}.
\eea
Also, in this case the re-heating and the nucleation temperature are related by 
\bea
T_R= T_n( 1+ \alpha(T_n))^{1/4}.
\eea
When calculating the frequency of the waves today, we use
\bea
f_0= \frac{a_*}{a_0} f_*.
\eea
Note that compared to the radiation case, in this case, since $a_*\gg a_n$, the redshift can become considerable. The vacuum domination regime is particularly relevant for our flat potential scenarios at high UV scales, where $\alpha$ can reach values of $\mathcal{O}(10^2-10^3)$. In these cases, the enhanced redshift between nucleation and reheating can shift the gravitational wave peak to lower frequencies, potentially bringing signals into the optimal sensitivity bands of space-based interferometers.
The temperature at which the transition takes place is given by the well known criterion of Eq.~ 22.124 of  \cite{Maggiore:2018sht}. In the next section we compare the distinctive features of the strong phase transitions, 
Fig.~\textcolor{blue}{10}, in comparison with medium and weak transitions, \fig{fig:GW_densityU1}-\fig{fig:GW_densitySU2not}.

\subsection{Parameter Space Analysis}

Our systematic scan over the coupling parameter space reveals several key phenomenological patterns. In Figures~\ref{fig:grid_GWU1}, \ref{fig:scansSU2U1_A}, \ref{fig:scansSU2U1_B}, and \ref{fig:scansSU2not}, we present comprehensive parameter space maps showing contours of constant $\alpha$ and $\beta/H_*$ values overlaid on the predicted gravitational wave energy density.

In Figs.~\ref{fig:grid_GWU1}--\ref{fig:scansSU2not} we have plotted planes $g$ vs $\lambda$ showing contours for specific values of $\alpha$ and $\beta/H_*$, while showing in the background the corresponding GW density $\Omega_{\rm{GW}} h^2$, up to the Big Bang Nucleosynthesis (BBN) bound \cite{Yeh:2022heq}.   Several important trends emerge from this analysis:
\begin{enumerate}
\item \textbf{Scale Dependence:} The UV cutoff $m_X$ acts as a crucial lever controlling transition strength. Moving from $m_X = 10^7$ GeV to $m_X = 10^9$ GeV increases $\alpha$ by 2-3 orders of magnitude while maintaining similar values of $\beta/H_*$.

\item \textbf{Coupling Sensitivity:} The gauge coupling $g$ exhibits a stronger influence on the transition parameters than that of the scalar self-coupling $\lambda$, reflecting the dominant role of the thermal gauge boson contributions in the barrier formation.

\item \textbf{Symmetry Breaking Pattern:} Complete symmetry breaking scenarios ($SU(2) \rightarrow \mathds{1}$) systematically produce stronger transitions than partial breaking ($SU(2) \rightarrow U(1)$ or $U(1) \rightarrow \mathds{1}$) due to the larger number of massive degrees of freedom.

\item \textbf{Fermion effects:} The inclusion of fermions generally weakens the transition strength by providing additional thermal pressure that opposes barrier formation, as seen in the comparison between Figures~\ref{fig:scan1SU2U1} and \ref{fig:scansSU2U1_B}.
\end{enumerate}

\begin{figure}[h]
    \centering
    \includegraphics[width=0.95\linewidth]{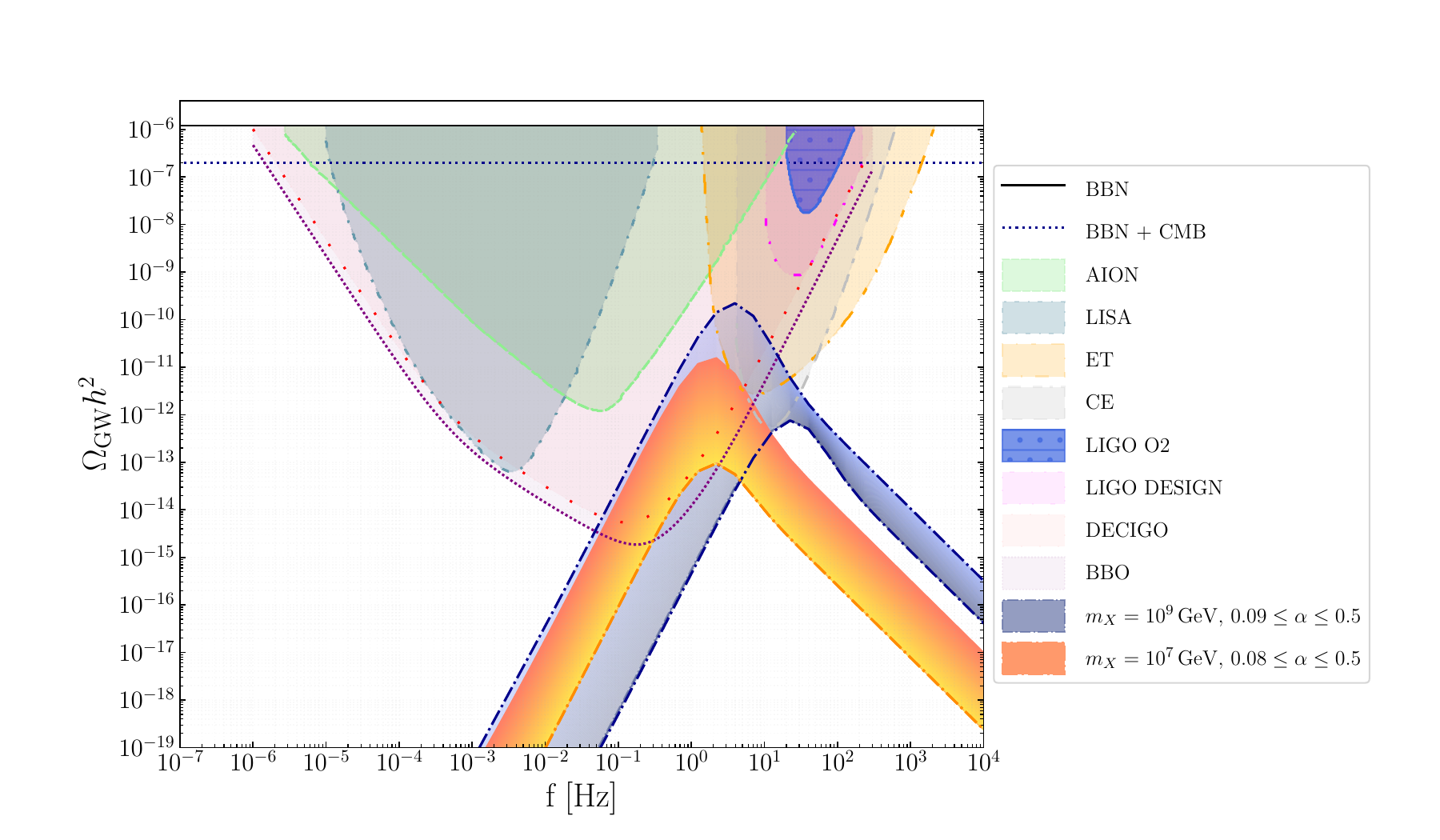}
    \caption{FOPT signatures of $U(1)$ models with $\alpha \leq 0.5$, where the lower bounds on the peak frequency, lie within experimental sensitivity. For $m_X=10^7$ GeV, the blue shaded area corresponds to $0.08 \leq \alpha \leq 0.5$,  and for $m_X=10^9$ GeV it is $0.09 \leq \alpha \leq 0.5$. }
    \label{fig:GW_densityU1}
\end{figure}

Then in Figs.~(\ref{fig:GW_densityU1}-\textcolor{blue}{11})
we show $\Omega_{\rm GW}(f)h^2$ for some values of $\alpha$, highlighting the regions that are accessible to future experiments, LISA, DECIGO, BBO, etc. 

\subsection{Gravitational Wave Spectral Analysis}

The gravitational wave spectra shown in Figures~\ref{fig:GW_densityU1}-\ref{fig:GW_densitySU2not} reveal the rich phenomenology accessible through flat potential scenarios. Each spectrum exhibits the characteristic broken power-law shape with a peak frequency determined by the Hubble scale at the transition epoch and the bubble separation scale. In \fig{fig:GW_densityU1}, we have plotted the FOPT signature of the $U(1)$ models for which we get a value of $\alpha\leq 0.5$ and a lower bound on the value of $\alpha$ that lies within the experimental sensitivity of an experiment shown in the plot. For $m_X=10^7$ GeV, the minimum sizable value is $\alpha=0.08$, while for $m_X=10^9$ GeV, $\alpha=0.09$. The near-independence of this threshold on the UV scale reflects the competing effects of enhanced transition strength (which increases the signal) and modified timing relationships (which can shift the peak frequency). 

The $U(1)$ breaking scenario exhibits several notable features:
\begin{itemize}
\item The spectral amplitude scales roughly as $\alpha^2$ in the weak transition regime, consistent with the expected quadratic dependence on the energy available for gravitational wave production.
\item Higher UV scales shift the peak to lower frequencies while increasing the amplitude, reflecting both the stronger transition and the enhanced redshift during vacuum domination.
\end{itemize}

 For the case of the FOPT signatures of the  $SU(2)\rightarrow U(1)$ model, we present the results in \fig{fig:GW_densitySU2U1}. For $m_X=10^9$ GeV, $n_f=0$ and $y=0$, we have plotted in shades of gray, densities corresponding to values of $\alpha$ up to $0.5$, where we can trust the simulations that produce the density profile that we use (see Appendix \ref{ap:FOPTandGW_nt}).  For $m_X=10^9$ GeV, $n_f=12$ and $y=0.35$, there is potentially a better chance to observe GW than in the previous case, since for $\alpha\leq 0.5$ there are regions that overlap DECIGO, BBO, ET and CE. For this case, we plotted $\alpha \leq 0.5$.

The partial $SU(2)$ breaking scenario demonstrates enhanced gravitational wave production compared to the $U(1)$ case due to the larger number of gauge degrees of freedom. The inclusion of fermions introduces several competing effects:
\begin{itemize}
\item Thermal fermion loops provide additional pressure opposing the phase transition, generally reducing $\alpha$.
\item However, the increased particle content also enhances the energy available for gravitational wave production once the transition occurs.
\item The Yukawa coupling strength $y$ allows fine-tuning of these competing effects, with moderate values ($y \sim 0.1-0.35$) providing optimal compromise between transition strength and detectability.
\end{itemize}
Finally, FOPT signature of the $SU(2)\rightarrow \mathds{1}$ model that lies in the sensitivity regions of various future experiments. Values smaller than $\alpha$ are of course possible. Also we have shown the case $\alpha=5$, to indicate the possible reach of this model. 

The complete $SU(2)$ breaking case represents the most optimistic scenario for gravitational wave detection, with three massive gauge bosons providing maximal thermal contribution to barrier formation. The resulting spectra consistently achieve the highest amplitudes across the parameter space we consider.

For very strong phase transitions, $\alpha \gg 1 $,  dominated by  by the vacuum energy, we follow \cite{Caprini:2024gyk}. In this case the GW, are mainly generated by the scalar field, since bubble walls interact weakly with the plasma and most energy goes into wall motion. Historically, GW production in this regime was estimated using the envelope approximation, which neglects contributions from regions where bubble walls have collided. However, recent studies and simulations indicate that these collided wall regions play a crucial role, as walls cannot lose momentum abruptly, making previous estimates of the stochastic GW background incomplete. For completeness, we present the template that we use in \eq{eq:strong_pt_temp}.

In \fig{fig:gw_grid_strong_pt_SU2U2} we present two examples of profiles for values of $\alpha$ within the LISA observation sensitivity that correspond to this case.
\begin{figure}[h]
    \centering
    \includegraphics[width=0.95\textwidth]{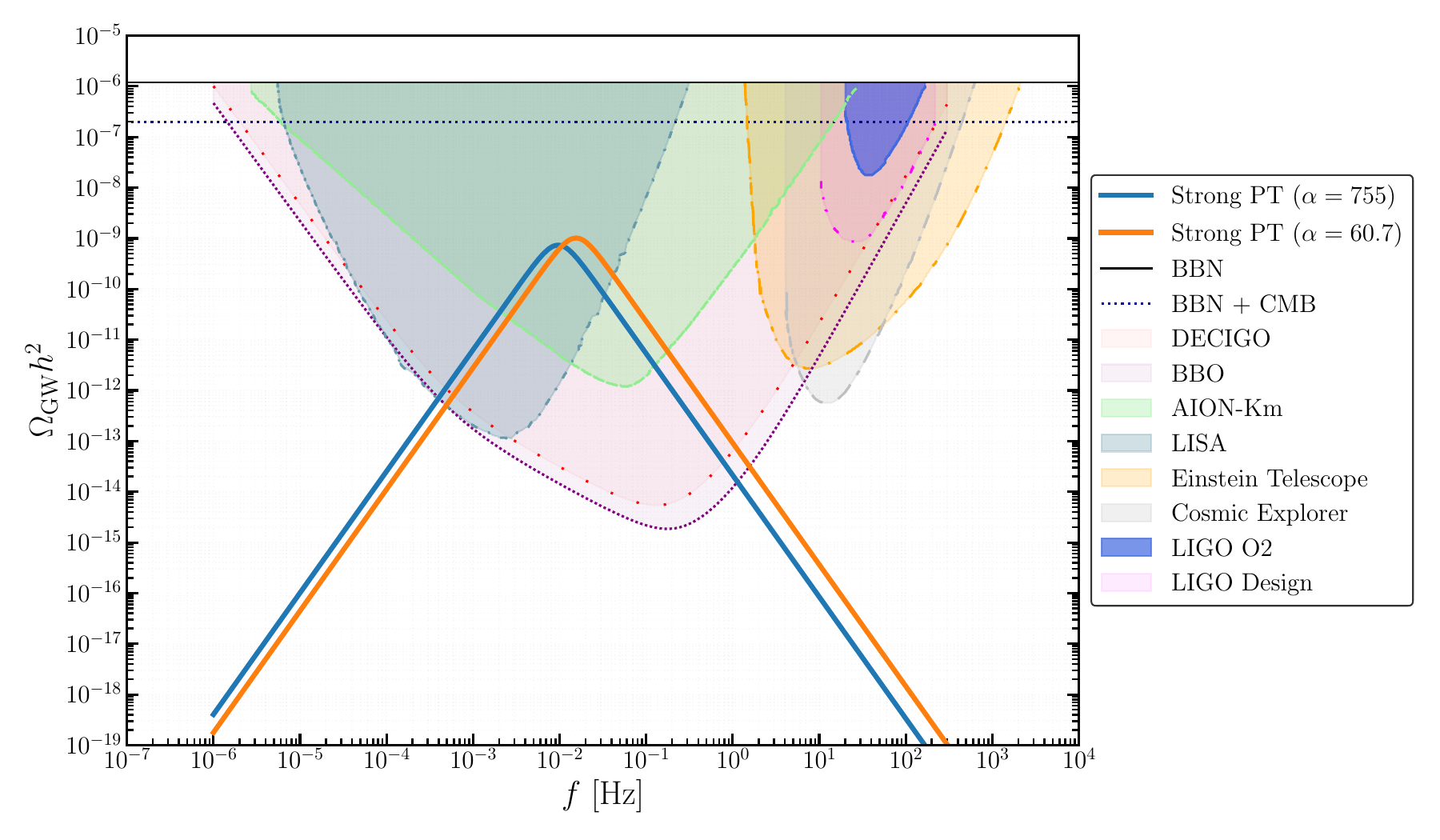}
    \caption{Example of the $\alpha\gg 1$ case, for some cases of $SU(2)\rightarrow U(1)$. The first case, $\alpha=755$, corresponds to the case of $m_X=10^9$ GeV, while the second case, $\alpha=60.7$ to the case of $m_X=10^{10}$ GeV. \label{fig:gw_grid_strong_pt_SU2U2}}
\end{figure}
The cases plotted in \fig{fig:gw_grid_strong_pt_SU2U2} correspond indeed to lines 5 and 8 of \tab{tbl:see_saw_ex}, we will talk about it in Section (\ref{eq:connection_neutr}). Note from \tab{tab:U(1)_params}-\tab{tab:SU(2)PB_params} that in each case there could be a case of vacuum domination ($\alpha\gg 1$). 

\begin{figure}
    \centering   \includegraphics[width=0.51\linewidth]{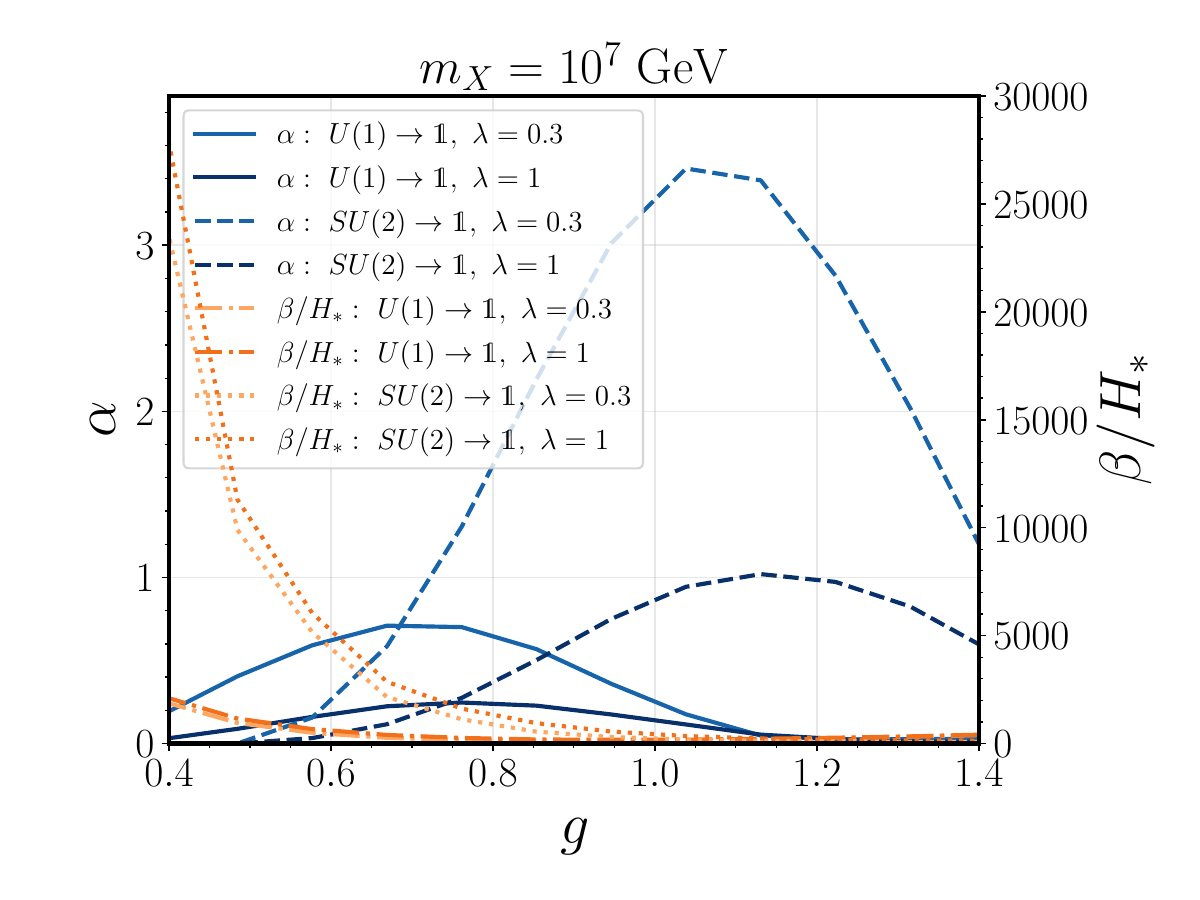}
    \hspace*{-0.93cm}
    \includegraphics[width=0.51\linewidth]{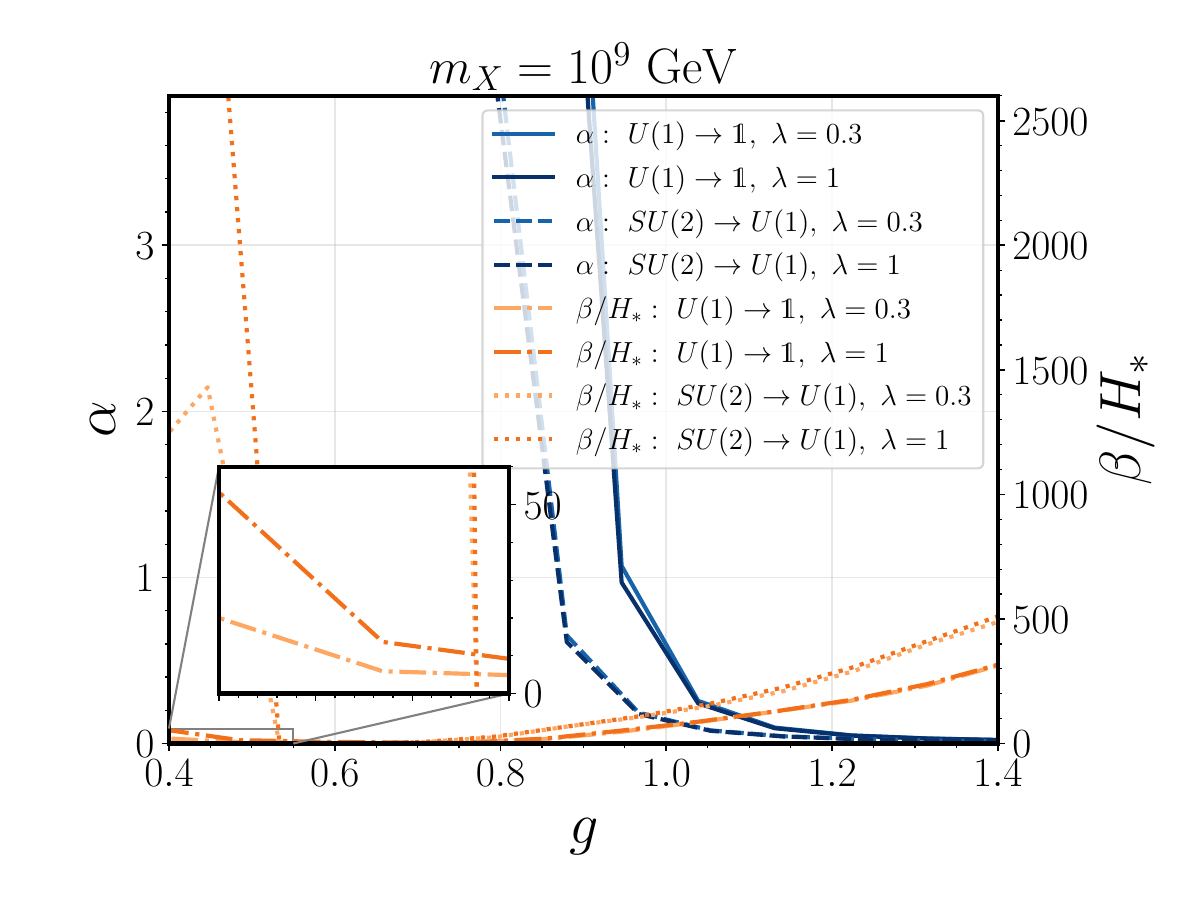}
    \caption{One-dimensional plots of $\alpha$ and $\beta$ versus $g$. \textbf{Left:} $m_X=10^7$ GeV, $n_f=0$, for $U(1)$ and the fully broken $SU(2)$ ($SU(2)\rightarrow \mathds{1}$). \textbf{Right:} $m_X=10^9$ GeV for $n=0$, for $U(1)$ and  $SU(2)\rightarrow U(1)$. A zoom highlights small $\beta/H_*$. $\alpha$ increases with the scale, consistent with the scaling behavior in Section~\ref{subsec:scaling}, and different $SU(2)$ breaking patterns are chosen to respect BBN bounds.\label{fig:alpha_beta_g_zoom}}    
\end{figure}
\begin{figure}
\centering
\includegraphics[width=0.51\linewidth]{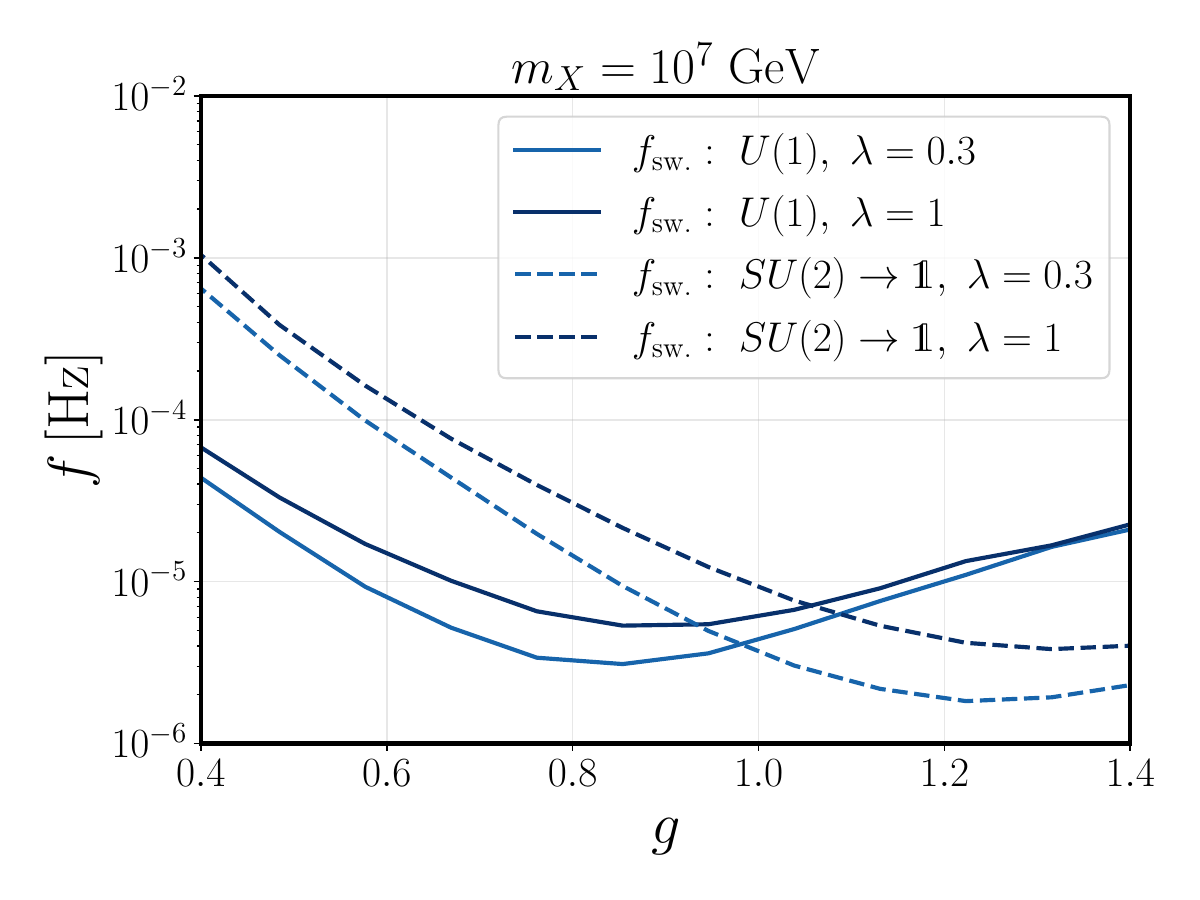}\\
\includegraphics[width=0.51\linewidth]{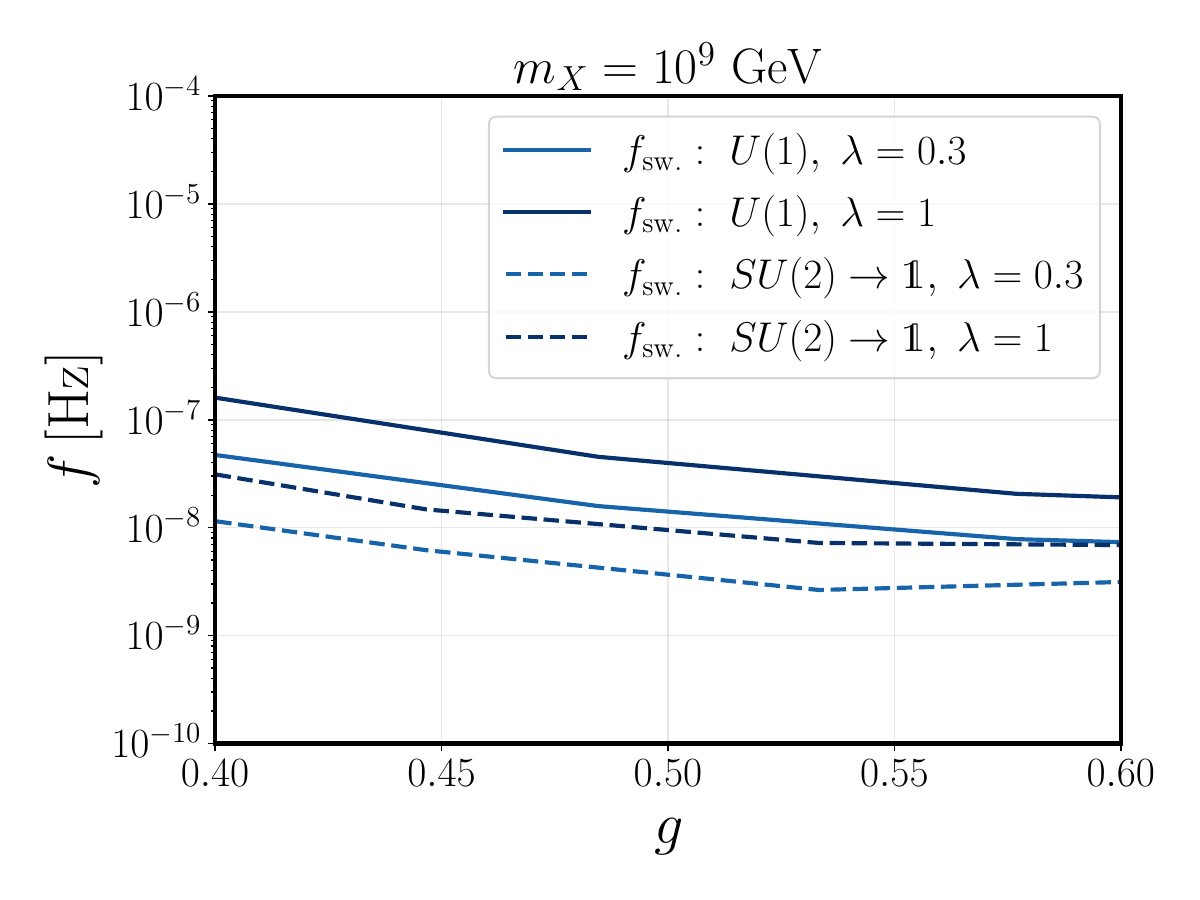}
   \hspace*{-0.66cm} \vspace{-0.5cm}
    \includegraphics[width=0.51\linewidth]
    {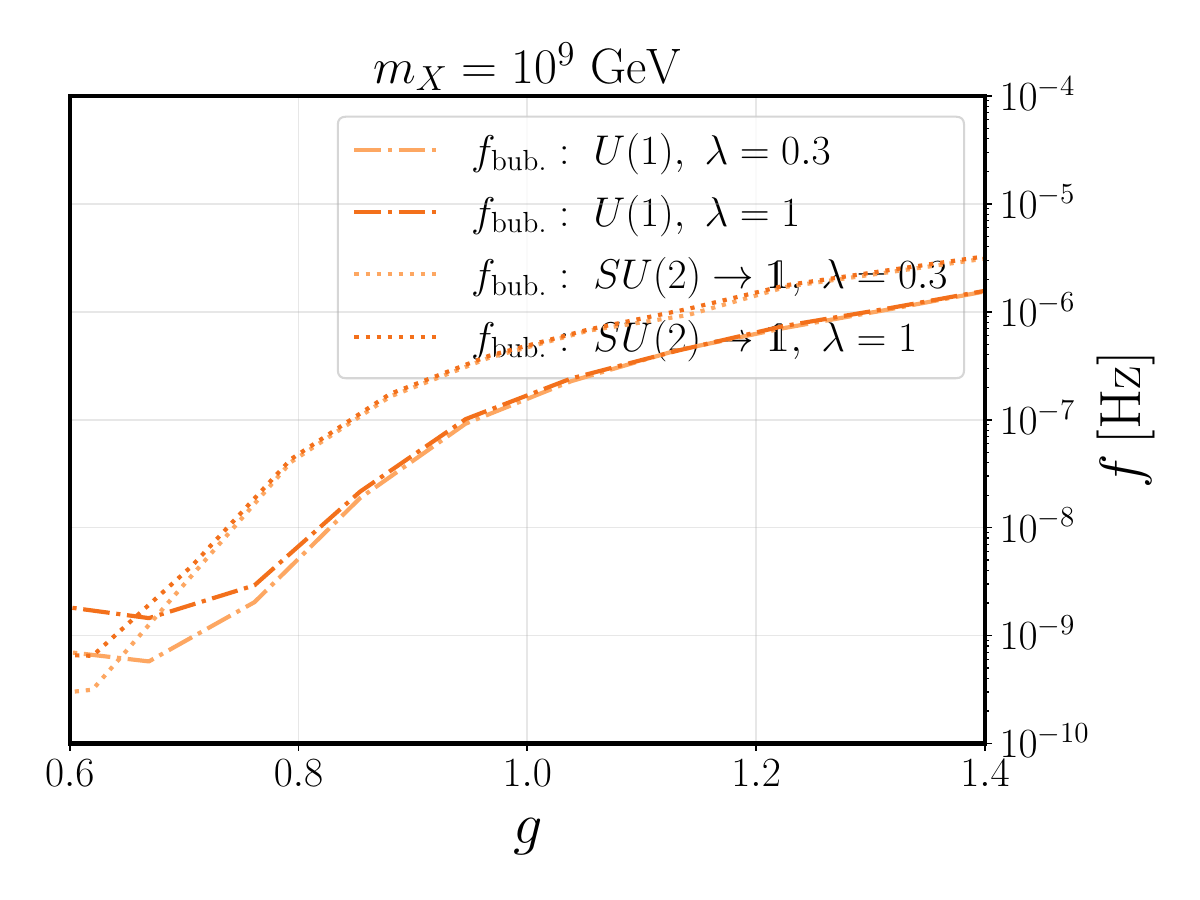}
    \caption{One-dimensional plots of the corresponding peak frequency for the cases presented in \fig{fig:alpha_beta_g_zoom}, for $m_X=10^7$ GeV (\textbf{top}) and for the case of $m_X=10^9$ GeV (\textbf{bottom}) . For $m_X=10^7$ GeV, all the plotted range of $g$ is dominated by sound waves, while for $m_X=10^9$ GeV, for $g<0.6$ the regime corresponds to sound waves domination and $g \geq 0.6 $ to bubble collisions domination.\label{fig:peak_freq_vs_g}}
\end{figure}
In \fig{fig:alpha_beta_g_zoom} we plot one dimensional plots showing the dependence of $\alpha$ and $\beta$ on $g$ for \textbf{(left)} $m_X=10^7$ GeV for $n_f=0$ for the $U(1)$ case and the maximally $SU(2)$ broken case, that is $SU(2)\rightarrow \mathds{1}$, and \textbf{(right)} $m_X=10^9$ GeV also for $n_f=0$. For this last case we have also plotted the case of $SU(2)\rightarrow U(1)$ (for $n_f=0$). Note the zoom insert that we have added such as to show the small values of $\beta/H_*$. For the $U(1)$ case, we can make a direct comparison in both cases. As the scale grows, the size of $\alpha$ increases significantly, as expected from the scaling behaviour we have described in Section \ref{subsec:scaling}. The reason we plot for one case $SU(2)\rightarrow \mathds{1}$ and for the other $SU(2)\rightarrow U(1)$ is because for the first case, a scale of $m_X=10^9$ GeV gives already values of $\alpha$ that saturate BBN bounds.
In \fig{fig:peak_freq_vs_g} we plot the peak frequency for the cases of \fig{fig:alpha_beta_g_zoom}. For the scale $m_X=10^7$ GeV, all the range considered for $g$ lies in the weak to medium $\alpha$ regime, and therefore the overall peak frequency of the signal corresponds to the sound wave frequency, $f_{\rsw}$, \eq{eq:fsw}. For the cases of $m_X=10^9$ GeV presented in \fig{fig:alpha_beta_g_zoom}, values of $g< 0.6$ correspond to the sound wave regime while $g\geq 0.6$ to the bubble collisions regime, \eq{eq:fbub}; for clarity, we displayed them in different figures. We did not plot $f_{\turb}$ as it is a subdominant contribution in both cases, and our intention was to plot the peak frequency.  We can see that the larger the gauge group, the greater the redshift (the peak frequency is lower for a fixed $g$). In these plots (\fig{fig:peak_freq_vs_g}-\fig{fig:alpha_beta_g_zoom}) we have plotted everything with $v_{\rsw}=1$ as this represents the upper limit. Other parameters entering into the computation of peak frequencies, see Appendix C, have been taken into account. There, we give the details of the values that we use for $\epsilon$ and the efficiency factor $\kappa_\nu$. 

\section{Discussion \label{sec:Discussion}}

The spontaneous breaking of the global lepton number symmetry considered in this work shares several dynamical features with other well-motivated extensions of the Standard Model. In particular, gauged $B-L$ models and frameworks with a Peccei–Quinn (PQ) \cite{Peccei:1977hh,Peccei:1977ur} symmetry also involve the vacuum expectation value of a new scalar field as the order parameter of the transition, potentially triggering a strong FOPT in the early Universe. However, our flat potential approach provides a more general framework that can accommodate various symmetry breaking scales without requiring specific model assumptions about gauge charges or axion couplings.

The key advantage of flat potential scenarios lies in their flexibility and naturalness. Unlike models with fine-tuned quartic couplings, the barrier formation in our approach arises predominantly from thermal effects, making the transition strength largely independent of the tree-level scalar self-interactions. This thermal dominance ensures robust predictions across wide parameter ranges and reduces sensitivity to specific UV completions of the effective theory.

While the specific scalar potentials and couplings differ, the thermodynamic parameters $\alpha$ and $\beta/H$ extracted in this work fall within ranges that can also arise in $B-L$ or PQ-breaking scenarios, suggesting that the gravitational wave (GW) templates derived here could be adapted to these related contexts. More precisely, the universal scaling $\alpha \propto (m_X/T_n)^4$ at high UV scales suggests that any effective field theory with similar higher-dimensional operators will exhibit comparable phenomenology, regardless of the underlying symmetry structure.

Our numerical analysis shows that for $\alpha \gtrsim 0.1$ and $\beta/H \lesssim 100$, the gravitational wave signal enters the projected sensitivity band of LISA, DECIGO, and BBO. In comparable $B-L$ or PQ-breaking models, phase transition parameters of this order are typically obtained for similar scalar quartic couplings and symmetry-breaking scales. This overlap in parameter space implies that a GW detection consistent with our predicted spectrum could not, by itself, uniquely identify the underlying symmetry, but could motivate a comparative study across multiple breaking patterns.

\subsection{Connection to Neutrino Masses \label{eq:connection_neutr}}
The scope of this work is not to present a complete model of neutrino masses but to show that it is possible within the context of a theory of neutrino mass generation to produce a potential of the type that we have been describing and such that the effective mass $M_R$ can be associated to the see-saw mechanism. To this end, in the context of  Type I seesaw, we have the well-known relation
\bea
\label{eq:see-saw_t1}
m_\nu \simeq \frac{y^2 v^2}{M_R}.
\eea
We note that, unlike the conventional Type-I seesaw realization where the
Standard Model Higgs vacuum expectation value explicitly enters the neutrino
mass formula, in our setup the dominant contribution to the Majorana mass
scale arises from the vacuum expectation value of the scalar field responsible
for spontaneous lepton-number breaking. As a result, the seesaw suppression
is controlled by the lepton-number breaking scale rather than directly by the electroweak Higgs vev.

If we want to reproduce the mass of the atmospheric neutrinos, we need $m_\nu \sim 0.05~{\rm eV}$ \cite{ParticleDataGroup:2024cfk}. In \tab{tbl:see_saw_ex}, we present representative benchmark points, covering the scales $10^{7}$ GeV to $\times10^{10}$ GeV that can satisfy the relation \eq{eq:see-saw_t1}, for one family of fermions within the model described in Section \ref{subsec:su2tu1}.

\begin{table}[h!]
\centering
\renewcommand{\arraystretch}{1.5}
\begin{tabular}{|c| c| c |}
\toprule
\hline
$v$ [GeV] & $y$ & $M_R$ [GeV]  \\
\hline
$3.5\times10^{5}$ & $6.39\times10^{-8}$ & $1\times10^{7}$      \\
$1.1\times10^{6}$ & $6.42\times10^{-8}$ & $1\times10^{8}$      \\
$3.6\times10^{6}$ & $6.2\times10^{-8}$ & $1\times10^{9}$      \\
$8.0\times10^{6}$ & $8.8\times10^{-8}$ & $1\times10^{10}$      \\
\hline
\end{tabular}
\caption{Representative benchmark points that satisfy $m_\nu=0.05\,$eV. \label{tbl:see_saw_ex}}
\end{table}
The Yukawa couplings displayed in Table~4 are correspondingly small. While
such values may appear to deviate from the usual naturalness motivation of
the seesaw mechanism, they are technically natural in the ’t~Hooft sense,
since setting them to zero enhances the symmetry of the theory. In this work,
we adopt these values as a consistent and minimal choice that allows us to
focus on the phase transition dynamics and the resulting gravitational-wave signals.

In this section, we would also like to connect the neutrino scale to strong first-order phase transitions, and therefore, we focus on two benchmark points that enter this regime in the observable LISA region.  In \fig{fig:gw_grid_strong_pt_SU2U2} we present the contour plots corresponding to lines 2 and 3 of \tab{tbl:see_saw_ex}. 
\begin{figure}[h]
    \centering
    \includegraphics[width=0.495\textwidth]{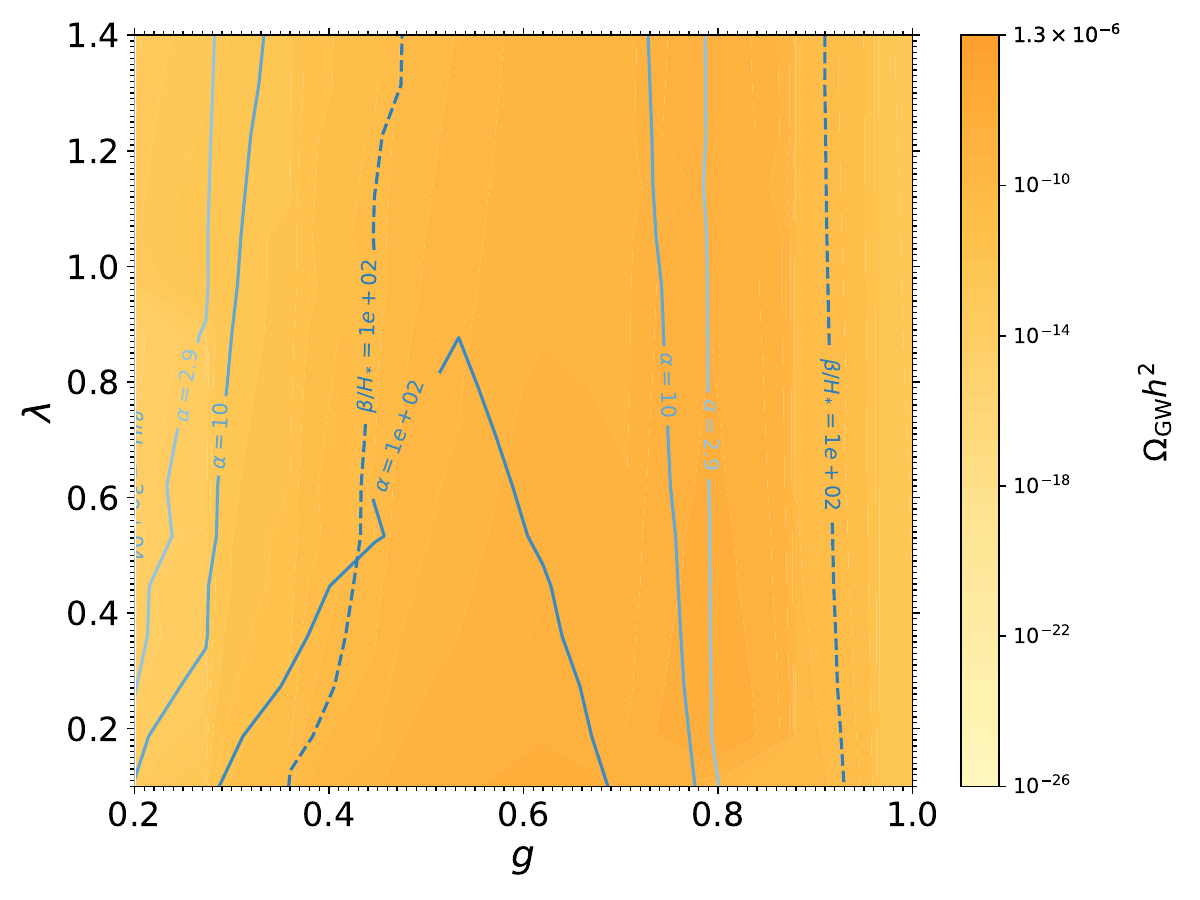}
\includegraphics[width=0.495\textwidth]{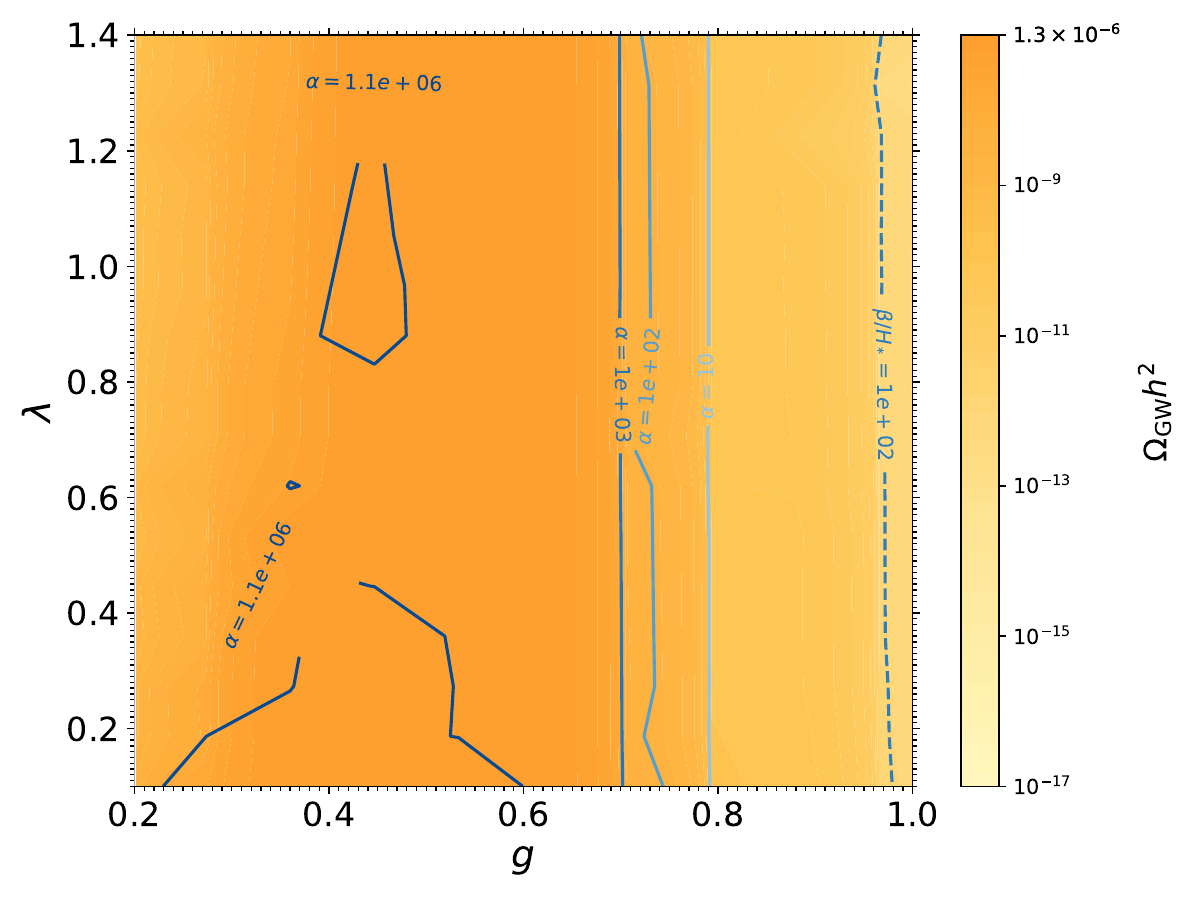}
\caption{For the $SU(2)\rightarrow U(1)$ case, we present two examples. \textbf{Left:} Second row of \tab{tbl:see_saw_ex} that is for a scale of $m_X=10^8$ GeV, and \textbf{Right:} Fourth row of \tab{tbl:see_saw_ex} that corresponds to a scale of $m_X=10^{10}$ GeV. }
    \label{fig:gw_grid_strong_pt_SU2U2}
    \end{figure}
%
As expected, since the value of the Yukawa coupling is so small, it does not differ greatly from the case of $y=0$ and $n_f=0$, \fig{fig:scanSU22U1}. This can be observed by comparing \fig{fig:scanSU22U1} (b) to \fig{fig:gw_grid_strong_pt_SU2U2}.

 If thermal inflation dilutes pre-existing asymmetry, leptogenesis must occur after the phase transition. This constrains $T_R > 10^9$ GeV for standard thermal leptogenesis, which may conflict with lower $m_X$ values, but in this case we have scenarios for which $T_R > 10^9$ GeV.

\subsection{Cosmological Implications Beyond Neutrino Masses}

Although our primary motivation centers on neutrino mass generation through the seesaw mechanism, the flat potential framework has broader cosmological implications that merit discussion.
From the perspective of baryogenesis, the parameter regions yielding the largest GW amplitudes correspond to transitions occurring at temperatures $T_* \sim 10^5$--$10^7$~GeV in our model. For $T_* \gtrsim 10^7$~GeV, the phase transition can produce even stronger gravitational wave signals. However, these scenarios typically involve very large values of the transition strength parameter, $\alpha \gg 1$, for which no dedicated numerical simulations currently exist to accurately model the resulting GW spectrum. Consequently, while such large-$\alpha$ regions are theoretically compelling, their predictions should be interpreted with caution.

In this high-temperature regime, the generation of heavy Majorana neutrino masses during the FOPT can significantly affect the efficiency of leptogenesis. In particular, the out-of-equilibrium dynamics driven by expanding bubble walls, combined with a shift in the timing of CP-violating decays, can alter the viable parameter space for generating the observed baryon asymmetry. A gravitational wave signal observed in this frequency range could therefore provide indirect evidence for a phase transition that played an active role in setting the matter–antimatter asymmetry of the Universe.

The thermal inflation context of our model introduces additional constraints on the cosmological history. If the transition occurs during a thermal inflation epoch, the subsequent reheating must be compatible with Big Bang nucleosynthesis and not overproduce gravitational waves beyond current CMB bounds. Our analysis shows that this requirement naturally selects UV scales $m_X \lesssim 10^{10}$ GeV for most symmetry breaking patterns, providing an upper bound on the effective theory cutoff.

Furthermore, the duration and strength of the phase transition affect the production of primordial black holes and other non-linear structures. Strong transitions with $\alpha \gg 1$ can generate significant density perturbations that seed black hole formation in specific mass ranges. The absence of such black holes in current observations could constrain the viable parameter space of flat potential models, complementing the direct gravitational wave limits.

\subsection{Multi-Messenger Observational Strategy}

The predicted gravitational wave signatures may be accompanied by complementary signals across multiple messengers and energy scales.
The scalar driving the transition can also leave collider and astrophysical imprints. For symmetry-breaking scales near the lower end of our range, Higgs–scalar mixing induces deviations in Higgs couplings, while additional scalar resonances could be accessible at future colliders. In particular, for UV scales $m_X \lesssim 10^8$ GeV, the mixing angle can reach $\sin\theta \sim 10^{-3}$, within the reach of precision Higgs measurements \cite{Djouadi:2017}, \cite{Cepeda:2019}.
At higher scales, the production of heavy right-handed neutrinos during the FOPT may contribute to the cosmic neutrino flux, with decay cascades producing ultra-high-energy neutrinos distinct from conventional sources. Next-generation observatories such as IceCube-Gen2 \cite{Lad:2023} could identify these via spectral hardness and flavor ratios.
The phase transition may also imprint the CMB by altering the sound horizon and damping acoustic oscillations. Though typically small, such effects could be probed by missions like CMB-S4 \cite{Abazajian:2019}, especially in combination with other cosmological data.
Altogether, collider experiments, astrophysical neutrino measurements, and space-based GW interferometers form a multi-messenger strategy to probe lepton number breaking in the early Universe.

\subsection{Cosmological History: Thermal Inflation Dynamics in Benchmark Scenarios}
The number of e-foldings in thermal  inflation can be quantified by noting that the era starts when $T\sim V_0^{1/4}\sim (m_0 m_X)^{1/2}$ and ends when $T\sim m_0$ \cite{Lyth:1995ka,Barreiro:1996dx}, therefore 
\bea
N \sim \frac{1}{2} \ln{\left(\frac{m_X}{m_0}\right)}.
\eea
Hence, we have $N=3.4, 4.6, 5.8, 6.9$ respectively for   $m_X=10^7,10^8,10^9,10^{10}$ GeV and $m_0=10^4$ GeV. Note that this estimation is independent of the value of $\alpha$. 



Regarding scalar and tensor perturbations generated during this brief
inflationary period, both are negligible compared to the primordial
fluctuations produced during the earlier high-scale inflationary epoch.
The amplitude of scalar perturbations generated during an inflationary
phase scales as $\Delta_{\mathcal{R}}^2 \sim (H/\dot{\phi})^2$, while tensor
perturbations scale as $\Delta_h^2 \sim (H/M_{\rm Pl})^2$. For the phase
transition considered here, with $H \sim H_* \sim 10^{-16} M_{\rm Pl}$
(corresponding to $T_* \sim 10^6$~GeV), these contributions are suppressed
by many orders of magnitude relative to the primordial signals and are
therefore observationally negligible, independently of the precise
duration of this vacuum-dominated phase.
The primary observational consequence of this vacuum-dominated phase is
the enhanced redshift of the gravitational wave signal from the phase
transition itself, as discussed in Section~\ref{sec:vac_dom}.

\subsection{Theoretical Challenges and Future Directions}

Several theoretical uncertainties affect the precision of our gravitational wave predictions and merit further investigation.

The most significant limitation concerns the treatment of strong phase transitions with $\alpha > 1$. Current numerical simulations of bubble dynamics and gravitational wave production are calibrated primarily for weak-to-moderate transitions, and their extrapolation to the vacuum domination regime introduces systematic uncertainties of order unity in the predicted amplitude. Improved simulations incorporating relativistic magnetohydrodynamics and non-linear gravitational effects are essential for robust predictions in this parameter range.

The calculation of bubble wall velocity represents another crucial source of uncertainty. While we employ state-of-the-art tools like 
\texttt{WallGo}~\cite{Ekstedt:2024fyq,vandeVis:2025plm}
for selected benchmarks (specifically for the $SU(2)\rightarrow \mathds{1}$ case), the systematic scan over parameter space relies on approximate analytical expressions that may not capture the full complexity of wall dynamics in flat potential scenarios. The interplay between friction from particle interactions and the driving pressure from the effective potential requires more sophisticated treatment, particularly near the transition between deflagration and detonation modes.

On the effective field theory side, our analysis assumes validity of the derivative expansion underlying the flat potential ansatz. For UV scales approaching the Planck mass, quantum gravitational corrections could become significant and modify the thermal barrier formation. Understanding these corrections requires insights from quantum gravity that are currently beyond our theoretical reach, but may become tractable through advances in holographic methods or asymptotic safety approaches.

Finally, the inclusion of gauge degrees of freedom beyond the minimal $U(1)$ and $SU(2)$ cases considered here opens rich phenomenological possibilities. Grand unified theories naturally provide multiple gauge groups that could undergo sequential breaking, each potentially generating gravitational wave signals at different frequencies. The interference and cross-correlation between these multiple signals could provide detailed information about the gauge structure and breaking pattern, motivating extended analysis beyond our current scope.



\section{Conclusions}

The detection of gravitational waves offers a promising new avenue to probe physics beyond the Standard Model, particularly in regimes inaccessible to current collider and laboratory experiments. In this work, we have studied the gravitational wave signatures of a first-order phase transition associated with the spontaneous breaking of lepton number symmetry, which dynamically generates Majorana masses for right-handed neutrinos. This mechanism is strongly motivated by the compelling evidence for neutrino masses and mixing, and it naturally arises in minimal seesaw frameworks capable of explaining the observed matter--antimatter asymmetry via leptogenesis. Our flat potential approach provides a particularly natural and robust framework for realizing strong first-order transitions without requiring fine-tuned scalar couplings or exotic particle content.  A striking outcome of the analysis is the emergence of a nontrivial scaling of the potential barrier, $V \propto m_0^{3} m_X$, revealing a distinctive and unexpected enhancement mechanism. 

We have analysed the finite-temperature effective potential governing the symmetry-breaking scalar sector, identified the conditions under which the transition is strongly first order, and computed the resulting parameters $\alpha$ (transition strength) and $\beta/H$ (inverse duration). These parameters determine the peak frequency and amplitude of the stochastic gravitational wave background, and their values are highly sensitive to the scalar couplings, vacuum expectation value, and thermal corrections from the particle content. Our systematic parameter space analysis reveals that the UV cutoff scale $m_X$ acts as the primary lever controlling transition strength, with values ranging from $10^7$ to $10^9$ GeV producing observable signals across complementary frequency ranges. This cut-off scale signals the lepton-number symmetry breaking scale and as such it is a window to probe one of the questions that the SM cannot answer.

In our study, we find broad regions of parameter space where the resulting gravitational wave spectrum is potentially observable, without violating current cosmological bounds from Big Bang nucleosynthesis, the cosmic microwave background, or large-scale structure. The robustness of these predictions stems from the thermal origin of barrier formation, which reduces sensitivity to specific UV completions and ensures reliable extrapolation across wide parameter ranges.

An important outcome of this work is the recognition that the interplay between thermal inflation and phase-transition dynamics can significantly alter the nucleation temperature, thereby shifting the gravitational wave peak frequency into the sensitivity bands of upcoming detectors. Moreover, we find that a strong first-order transition is achievable in well-motivated regions of parameter space without requiring extreme scalar couplings, particularly when thermal loop effects are included in the analysis. 

The correlation between neutrino mass scales and gravitational wave observables provides a unique bridge connecting high-energy particle physics with precision cosmological measurements. This also provides a novel indirect test of seesaw-type models, offering complementary constraints to those from oscillation data, neutrinoless double beta decay, and collider searches.

This multi-messenger approach is particularly powerful because it probes the same underlying physics - lepton number violation and right-handed neutrino mass generation - through completely independent observational channels.

The predicted signals lie in the milli-Hertz to deci-Hertz frequency range, making them ideal targets for future space-based interferometers such as LISA~\cite{LISA:2017pwj}, DECIGO~\cite{Kawamura:2011zz}, and BBO~\cite{Crowder:2005nr}. Our analysis shows that different symmetry breaking patterns produce characteristic spectral signatures that could enable discrimination between $U(1)$, partial $SU(2)$, and complete $SU(2)$ breaking scenarios through their distinct peak frequencies and amplitudes.

Looking ahead, a positive detection of such a gravitational wave background would have profound implications for both particle physics and cosmology. It would point to the existence of new scalar fields and symmetries in the early universe, confirm the cosmological origin of right-handed neutrino masses, and provide an unprecedented experimental handle on physics at energy scales far beyond those of terrestrial accelerators. Beyond the immediate discovery potential, such observations would establish gravitational wave astronomy as a precision tool for probing the detailed dynamics of phase transitions, potentially revealing subtle features of bubble nucleation, wall velocity evolution, and hydrodynamic instabilities that are currently beyond theoretical reach.

Even in the absence of a signal, the constraints derived from null results in the relevant frequency bands would exclude large classes of lepton-number-breaking models and refine the viable parameter space for neutrino mass generation. Furthermore, the methodology developed here can be naturally extended to other symmetry-breaking scenarios, such as multi-step transitions, hidden sector gauge groups, or scalar sectors with non-trivial vacuum structures, each of which could yield distinctive gravitational wave signatures. The systematic approach we have developed - combining effective field theory methods with thermal field theory calculations and gravitational wave phenomenology - provides a template for analyzing phase transitions in broader classes of beyond-Standard Model scenarios.

The coming decades, with the operation of LISA, DECIGO, and BBO, promise to turn gravitational wave astronomy into a precision probe of high-energy particle physics, opening a new observational frontier for the study of fundamental symmetries and the origin of mass. Our work demonstrates that flat potential scenarios offer particularly promising targets for this exciting program, with observable signals achievable across natural parameter ranges and robust theoretical predictions that are largely independent of specific UV completions. The convergence of gravitational wave observations, precision neutrino measurements, and high-energy collider data in the coming decade may finally resolve the longstanding puzzle of neutrino mass generation while revealing new aspects of the thermal history of the early universe.

\section{Acknowledgments}

We thank J\"orn Kersten, Injun Jeong, Seongchan Park, Stefano Scopel and Juhoon Son for useful
discussions. We also thank Jorinde Van de Vis and Philipp Schicho for help with \texttt{WallGo}.
The work of L.V.S.\ is supported by the National
Research Foundation of Korea (NRF) grant no.\ RS-2023-00273508.
G. B. is supported by the Spanish grants  CIPROM/2021/054 (Generalitat Valenciana), PID2020-113775GB-I00 (AEI/10.13039/501100011033), and by the European ITN project HIDDeN (H2020-MSCA-ITN-2019/860881-HIDDeN). Y. P is supported by IBS under the project code IBS-R018-D1.

\appendix
\section{Integrating out heavy fields \label{app:heavyfields}}
Suppose that at a particular scale of breaking,  there is a heavy real scalar $X$, with mass $M_X$, and a complex singlet $\Phi$. The UV potential  is
\begin{equation}
    V_{\rm UV} = \tfrac12 M_X^2 X^2 + \big(\mu\, X \Phi^3 + \mu^\ast X \Phi^{\dagger 3}\big) + V_{\rm other}(\Phi),
\end{equation}
where $V_{\rm other}(\Phi)$ contains mass terms for $\Phi$, and we assume $|\Phi|^4$ is set to zero at tree level by choice of UV parameters or classical scale invariance. Notice that no coupling $X|\Phi|^2$ is present in the UV by construction. We can solve the classical EOM for $X$, treating $X$ as heavy in order to integrating it out:
\begin{equation}
\frac{\partial V}{\partial X}=M_X^2 X + \mu \Phi^3 + \mu^\ast \Phi^{\dagger 3} + \cdots =0
\quad\Rightarrow\quad
X \simeq -\frac{\mu}{M_X^2} \Phi^3 - \frac{\mu^\ast}{M_X^2} \Phi^{\dagger 3} + \mathcal{O}\!\big(|\Phi|^5/M_X^4\big).
\end{equation}
Inserting back into $V$, we obtain:
\begin{equation}
V_{\rm Eff. } \supset \tfrac12 M_X^2 X^2 \simeq
\tfrac12 M_X^2\left(\frac{\mu}{M_X^2}\Phi^3 + \frac{\mu^\ast}{M_X^2}\Phi^{\dagger 3}\right)^2 .
\end{equation}
Finally, expanding, we get
\bea
V_{\rm Eff. } \supset -\frac{\mu^2}{2M_X^2} \Phi^6 -\frac{(\mu^\ast)^2}{2M_X^2} \Phi^{\dagger 6}
- \frac{|\mu|^2}{M_X^2}\,|\Phi|^6.
\eea
Hence the real part includes a $|\Phi|^6$ term:
\begin{equation}
V_{\rm Eff. }\supset \lambda_6\,|\Phi|^6 \quad\text{with}\quad
\lambda_6 \simeq -\frac{|\mu|^2}{M_X^2}\,.
\end{equation}

\section{Further contributions to the thermal potential \label{app:furthercontThPot}}

\subsection*{Thermal functions}
The thermal masses contribute to the thermal potential through the argument of the free energy, as they shift 
the field-dependent masses of the associated particles, in the thermal potential of Eq.~(2.2) where the thermal functions are defined as
\begin{align}
\label{eq:JBJF}
J_B(y^2) = \Re \int_0^\infty dx \, x^2 \ln\left[1 - e^{-\sqrt{y^2+x^2}}\right], \\
J_F(y^2) = \Re \int_0^\infty dx \, x^2 \ln\left[1 + e^{-\sqrt{y^2+x^2}}\right], 
\end{align}
which in the high-temperature expansion are given by
\begin{align}
\label{eq:JBJF_exp}
J_B(y^2) \simeq -\frac{\pi^4}{45} + \frac{\pi^2}{12}y^2 - \frac{\pi}{6}y^3 - \frac{y^4}{32}\ln\frac{y^2}{a_B} + \ldots, \\
J_F(y^2) \simeq \frac{7\pi^4}{360} - \frac{\pi^2}{24}y^2 - \frac{y^4}{32}\ln\frac{y^2}{a_F} + \ldots, \quad y = m/T, 
\end{align}
with $a_B = 16\pi^2 e^{3/2-2\gamma_E}$ and $a_F = \pi^2 e^{3/2-2\gamma_E}$, where $\gamma_E$ is the 
Euler–Mascheroni constant. When temperature corrections are important, the effective squared mass entering the thermal functions is shifted by the self-energy correction
\[m_{\text{Eff. },p}^2(\phi, T) = m_p^2(\phi) + \Pi_p(T). \]

\subsection*{Leading-order thermal masses}

Our numerical implementation evaluates thermal masses (propagators) for different gauge group breaking patterns. 
For completeness, we provide the leading-order formulas for the Debye (gauge) and thermal (scalar) masses 
in the hard thermal loop expansion for a $SU(N)$ theory:
\begin{align}
\label{eq:Props}
\Pi_g(T) \simeq g^2 T^2\left(\frac{C_{2A}}{3} + \sum_f k_f\frac{T_{R_f}}{6} + \sum_s \frac{T_{R_s}}{6}\right),\\
\Pi_\phi^{\text{(gauge)}}(T) \simeq \frac{g^2 C_{2R}}{4}T^2, \quad \Pi_\phi^{(\lambda)}(T) = 
\frac{5}{128}\frac{\lambda}{m_X^2}T^4,
\end{align}
where $C_{2A}$ is the quadratic Casimir of the adjoint representation (which is equal to the Dynkin index of the adjoint representation), $T_{R_{f,s}}$ is the Dynkin index of representation $R_{f,s}$, for a given  fermion $f$ or scalar $s$, and $C_{2R}$ is the quadratic Casimir of representation $R$. The factor $k_f=1$ for Dirac fermions while $k_f=1/2$ for Majorana fermions or Weyl fermions. Note that the gauge propagator, $\Pi_g$, and the contribution from the gauge coupling to the scalar propagator, $\Pi_\phi$, are exactly like the $\lambda \phi^4$ theory (see  \cite{Braaten:1990mz} and \cite{Rebhan:1993az} for original computations) since at this level they are not affected by the quartic coupling. For $U(1)$ case, the expression \eq{eq:Props} becomes  $\Pi_g(T) \simeq g^2 T^2\left(  \frac{1}{6} \sum_{f} c_f^2+ \frac{1}{6} \sum_{s} c_s^2\right)$, where the sum is over fermions, $f$ and scalars, $s$, and  $c_f$ and $c_s$ the corresponding charges. Note that there is not an adjoint representation in U(1). Also for the $U(1)$ case, $\Pi_\phi^{\text{(gauge)}}(T) \simeq \frac{g^2 c_\phi^2}{4}T^2$, where $c_\phi$ is the $U(1)$ charge of $\phi$.
The dimension-six 
contribution $\Pi_\phi^{(\lambda)}(T)$ arises from the $\phi^6$ operator as derived below. When fermions are included, they contribute additional thermal corrections:
\bea
\Pi_\phi^{\text{(Yuk)}}(T) \simeq \begin{cases}
\frac{y^2}{12}T^2, & \text{Dirac fermion},\\
\frac{y^2}{24}T^2, & \text{Majorana or Weyl fermion}.
\end{cases} 
\eea

\subsection*{Specific cases}

\subsubsection*{$U(1)$ without fermions (for an effective scalar real degree of freedom)}
\bea
\Pi_g(T) \simeq \frac{g^2}{6}T^2, \quad \Pi_\phi (T) \simeq \frac{g^2}{4}T^2 + \frac{5}{128}\frac{\lambda}{m_X^2}T^4, 
\eea
with  $c_\phi=1$.
\subsubsection*{$U(1)$ with $N_f$ Majorana fermions (for an effective scalar real degree of freedom)}
\begin{align}
\Pi_g (T)&\simeq N_f c_\phi^2 \frac{g^2}{6}T^2 + \sum_\psi\frac{c^2_\psi g^2}{6}T^2,\\
\Pi_\phi (T)&\simeq \frac{g^2}{4}T^2 + \frac{5}{128}\frac{\lambda}{m_X^2}T^4 + \sum_\psi \frac{ y_\psi^2}{24}T^2.
\end{align}
For the case we considered in Section (\ref{sec:u1}) we have two Majorana fermion flavors with charges  $|c_\psi| = 1/2$ and $c_\phi=1$. Therefore, in this case, 
\begin{align}
\Pi_g(T) \simeq  \frac{1}{24} N_f g^2 T^2 +\frac{1}{6} g^2 T^2
\end{align}
\subsubsection*{$SU(2) \rightarrow U(1)$ with an adjoint scalar}

Without fermions, we have: 
\bea
\Pi_{g}(T) \simeq g^2 T^2, \quad \Pi_\phi(T) \simeq \frac{g^2}{2}T^2 + \frac{5}{128}\frac{\lambda}{m_X^2}T^4, 
\eea
with $C_{2A}=2$.

With fermions (in the adjoint representation):
\bea
\Pi_{g}(T) \simeq \frac{3}{2} g^2 T^2, \quad \Pi_\phi(T) \simeq \frac{g^2}{2}T^2 + \frac{5}{128}\frac{\lambda}{m_X^2}T^4  + 
\sum_\psi \frac{y_\psi^2}{24}T^2,
\eea
with $C_{2A}=2$.

\subsubsection*{$SU(2) \rightarrow \mathds{1}$ with a scalar in the fundamental representation}
\bea
\Pi_{g}(T) \simeq \frac{3}{4}g^2 T^2, \quad \Pi_\phi(T) \simeq \frac{3}{16}g^2 T^2 + 
\frac{5}{128}\frac{\lambda}{m_X^2}T^4,
\eea
with $C_{2A}=2$,  $T_R=1/2$, $C_{2R}=3/4$.

\bigskip
\subsection*{Propagator for the $\phi^6$ term}

For the U(1) case, we consider the interaction of a scalar field \( \phi \) with the U(1) gauge boson \( A_\mu \), described by the interaction term
\begin{align}
    \mathcal{L} \supset \frac{1}{2} g^2\phi^2 A_\mu A^\mu,
\end{align}
where \( g \) is the U(1) gauge coupling. This yields a field-dependent mass \( m_A^2(\phi) = g^2 \phi^2 \) for the gauge boson. The relevant leading-order thermal self-energies contributing to the Daisy potential can be approximated by
\begin{align}
    \Pi_\phi &= T \Sumint_n \! \frac{d^3 k }{(2 \pi)^3}\left\{ \frac{g^2 g^{\mu\nu} \left(g_{\mu\nu} - \frac{p_\mu p_\nu}{m_A^2} \right)}{\omega_n^2 + k^2 +m_A^2} + \Sumint_m \! \frac{d^3 p }{(2 \pi)^3}\frac{\lambda}{8m_X^2}\frac{45}{(\omega_n^2 + k^2 +m_\phi^2)(\omega_m^2 + p^2 +m_\phi^2)} \right\} \\
    &\quad \approx \frac{g^2 T^2}{4} + \frac{5}{128} \frac{\lambda}{m_X^2}T^4 \\
    \Pi_g &= g^2 T \Sumint_n \frac{d^3 k }{(2 \pi)^3} \frac{3}{\omega_n^2 + k^2 +m_\phi^2} \\
    &\quad \approx \frac{g^2 T^2}{6}.
    \label{eq:self-energy}
\end{align}

These self-energies are derived from one-loop diagrams at finite temperature using the imaginary-time formalism, where the sum over Matsubara frequencies $\omega_n = 2\pi n T$ replaces the usual energy integral. For the scalar self-energy, the two-loop ``figure-8" diagram arises from the higher-dimensional interaction
\begin{align}
V_\text{int} \supset \frac{\lambda}{8 m_X^2} \phi^6.
\end{align}
This operator leads to a factorized diagram in which two of the six fields are treated as external legs, while the remaining four form two internal scalar loops.

The combinatorics proceed as follows: there are \( \binom{6}{2} = 15 \) ways to select two external lines, and the remaining four internal legs can be contracted into two indistinguishable pairs in \( \frac{1}{2!} \binom{4}{2} = 3 \) ways. Thus, the full diagrammatic factor is \( 15 \times 3 = 45 \), giving an effective vertex of
\begin{align}
\frac{45 \lambda}{8 m_X^2}.
\end{align}
Combining this with the thermal loop integrals, the two-loop contribution to the scalar self-energy becomes
\begin{align}
    \Pi_\phi^{(\phi^6)} 
    &= \frac{45 \lambda}{8 m_X^2} \times \frac{T^4}{144}
    = \frac{5 \lambda}{128 m_X^2} T^4.
\end{align}

\begin{figure}
\begin{minipage}{0.32\textwidth}
\centering
\begin{tikzpicture}
  \begin{feynman}
    \vertex (L) at (-1.7,0);
    \vertex (R) at (1.7,0);
    \vertex (C) at (0,0);
    \vertex (T) at (0,1.7);

    \diagram* {
      (C) -- [boson, half left, looseness=1.6, ->] (T) -- [boson, half left, looseness=1.6, ->] (C),
      (L) -- [scalar, ->] (C) -- [scalar, ->] (R),
    };
  \end{feynman}
\end{tikzpicture}\\
\vspace*{0.2cm}
(a)
\end{minipage}
\begin{minipage}{0.32\textwidth}
\centering
\begin{tikzpicture}
  \begin{feynman}
    \vertex (L) at (-2,0); \vertex (C) at (0,0); \vertex (R) at (2,0);
    \vertex (T1) at (0,1.5); \vertex (B1) at (0,-1.5);

    \diagram* {
      (C) -- [scalar, half left, looseness=1.6, ->] (T1) -- [scalar, half left, looseness=1.6, ->] (C),
      (C) -- [scalar, half right, looseness=1.6, ->] (B1) -- [scalar, half right, looseness=1.6, ->] (C),
      (L) -- [scalar, ->] (C) -- [scalar] (R),
    };
  \end{feynman}
\end{tikzpicture}\\
\vspace*{0.2cm}
(b)
\end{minipage}
\begin{minipage}{0.32\textwidth}
\centering
\begin{tikzpicture}
  \begin{feynman}
    \vertex (L) at (-1.7,0); \vertex (R) at (1.7,0); \vertex (C) at (0,0);
    \vertex (T) at (0,1.7);

    \diagram* {
      (C) -- [scalar, half left, looseness=1.6, ->] (T) -- [scalar, half left, looseness=1.6, ->] (C),
      (L) -- [boson, ->] (C) -- [boson, ->] (R),
    };
  \end{feynman}
\end{tikzpicture}\\
\vspace*{0.2cm}
(c)
\end{minipage}
\caption{Finite-temperature self-energy diagrams relevant for the calculation of $\Pi_\phi$ and $\Pi_g$ in \eq{eq:self-energy}. \label{fig:Finite_Tcorr}}
\end{figure}

Thus, the total leading-order thermal self-energy for the scalar field including both gauge boson and $\phi^6$ contributions is
\begin{align}
    \Pi_\phi(T) \approx \frac{g^2 T^2}{4} + \frac{5}{128} \frac{\lambda}{m_X^2} T^4,
\end{align}
while the gauge boson thermal self-energy is
\begin{align}
    \Pi_g(T) \approx \frac{g^2 T^2}{6}.
\end{align}
In the \( \lambda \phi^4 \) theory, infrared divergences due to bosonic zero modes are regulated by the ring (daisy) resummation \cite{Carrington:1991hz, Arnold:1992rz}, leading to an improved thermal potential:
\begin{align}
    V(\phi, T) = V_\text{Tree}(\phi) + V_{1\text{-loop}}^{T=0}(\phi) + V_{1\text{-loop}}^{T}(\phi, T) + V_\text{daisy}(\phi, T),
\end{align}
with the Daisy term
\begin{align}
\label{eq:VDaisy}
    V_{\rm Daisy}(\phi, T) = - \sum_i \frac{n_i T}{12 \pi} \left[ \left(m_i^2(\phi) + \Pi_i(T)\right)^{3/2} - m_i^3(\phi) \right].
\end{align}
This procedure remains valid in the presence of higher-dimensional operators like \( \phi^6 \), provided the temperature remains below the EFT cutoff (\( T \ll m_X \)) and the dimension-six coupling \( \lambda / m_X^2 \) is perturbative. In such cases, the dominant thermal correction arising from the $\phi^6$ operator scales as \( T^4/m_X^2 \), and can be included consistently in the thermal mass entering the Daisy resummation. {For the $SU(2)$ cases, the derivation is similar and yields exactly the same result.

\section{Formulas for GW density \label{ap:FOPTandGW_nt}}

As is well established, the energy density of the stochastic gravitational-wave
background produced during a first-order phase transition is determined by a set
of macroscopic parameters, namely the nucleation temperature $T_n$ (or
equivalently the plasma temperature $T_*$
\footnote{$T_n \approx T_*$ for weak to intermediate transitions, while for
vacuum-dominated transitions they follow the hierarchy of
\eq{eq:temphierarchies}.}), the phase-transition strength $\alpha$, the inverse
duration parameter $\beta$, the sound speed $c_s$, the bubble-wall velocity $v_w$,
and the efficiency factors $\kappa_\nu$ and $\epsilon$. In terms of these
quantities, one can employ standard expressions for the gravitational-wave
contributions from sound waves and magnetohydrodynamic turbulence, as derived in
Refs.~\cite{Hindmarsh:2013xza,Hindmarsh:2015qta} and
\cite{Caprini:2009yp,Binetruy:2012ze}, respectively. The sound-wave contribution is
given by
\bea
\Omega_{\rsw} h^2(f)=2.65 \times 10^{-6}(H_*\tau_{\rsw})
\left(\frac{H_*}{\beta}\right) v_w
\left(\frac{\kappa_\nu \alpha }{1+\alpha }\right)^2
\left(\frac{g_*}{100}\right)^{-1/3}
\left(\frac{f}{f_{\rsw}}\right)^3
\left(\frac{7}{4+3 \left(f/f_{\rsw}\right)^2}\right)^{7/2}\hspace*{-0.6cm},
\eea
where
$\tau_{\rsw}=\min\!\left[H_*^{-1},\,R_*/\bar{U}_f\right]$
characterizes the effective duration of the sound-wave source
\cite{Ellis:2019oqb}. Here $R_*$ denotes the mean bubble separation, satisfying
$H_*R_* = v_w(8\pi)^{1/3}(\beta/H_*)^{-1}$, and the root-mean-square fluid velocity
can be approximated as
\bea
\bar{U}_f^2 \simeq \frac{3}{4}\left(\frac{\kappa_\nu\alpha}{1+\alpha}\right).
\eea
The efficiency factor $\kappa_\nu$ is approximated by \cite{Espinosa:2010hh}
\footnote{\texttt{WallGo} defines the efficiency factor as
$\kappa_{\rm WallGo}=E_{\rm kin}/\Delta\rho_{\rm vac}$, which is related to
$(\kappa_\nu\alpha)/(1+\alpha)=\kappa_{\rm WallGo}\alpha/(1+\alpha)$.
Throughout this work we use the \texttt{WallGo} definition, which agrees well with
\eq{eq:kappanuapprox} over most of the parameter space we explore.}
\bea
\label{eq:kappanuapprox}
\kappa_\nu \simeq
\begin{cases}
\alpha\left(0.73+0.83\sqrt{\alpha}+\alpha\right)^{-1},
& v_w \sim 1, \\
6.9\,v_w^{6/5}\alpha
\left(1.36-0.037\sqrt{\alpha}+\alpha\right)^{-1},
& v_w \ll 1 .
\end{cases}
\eea
The corresponding peak frequency is
\bea
\label{eq:fsw}
f_{\rsw}=1.9\times 10^{-5}\,\frac{1}{v_w}
\left(\frac{\beta}{H_*}\right)
\left(\frac{T_*}{100\,{\rm GeV}}\right)
\left(\frac{g_*}{100}\right)^{1/6}
{\rm Hz},
\eea
where $g_*$ denotes the number of relativistic degrees of freedom.

\medskip
On the other hand, magnetohydrodynamic turbulence in the plasma provides a
subleading contribution to the gravitational-wave signal, with energy density
\bea
\Omega_{\turb} h^2(f)=3.35 \times 10^{-4}
\left(\frac{\beta}{H_*}\right)^{-1}
\left(\frac{\epsilon\,\kappa_\nu \alpha}{1+\alpha}\right)^{3/2}
\left(\frac{g_*}{100}\right)^{-1/3} v_w
\frac{(f/f_{\turb})^3(1+f/f_{\turb})^{-11/3}}
{1+8\pi f/h_{*,0}} ,
\eea
where the redshift factor to today is
\bea
\label{eq:h0st}
h_{*,0}=16.5
\left(\frac{T_*}{10^8\,{\rm GeV}}\right)
\left(\frac{g_*}{100}\right)^{1/6}
{\rm Hz}.
\eea
We assume that the turbulence efficiency factor can be written as
$\kappa_{\rm turb}=\epsilon\,\kappa_\nu$, where $\epsilon$ parametrizes the
fraction of bulk fluid motion converted into turbulence. Numerical simulations of
magnetohydrodynamic turbulence~\cite{Hindmarsh:2015qta} indicate that at most
$5\%$--$10\%$ of the kinetic energy in bulk flows is transformed into vorticity.
We therefore adopt the conservative choice $\epsilon=0.05$
\cite{Caprini:2015zlo}, which we employ throughout this work.
The peak frequency is
\bea
\label{eq:fturb}
f_{\turb}=2.7\times 10^{-5}\,\frac{1}{v_w}
\left(\frac{\beta}{H_*}\right)
\left(\frac{T_*}{100\,{\rm GeV}}\right)
\left(\frac{g_*}{100}\right)^{1/6}
{\rm Hz}.
\eea

\medskip
Ref.~\cite{Caprini:2024gyk} recently proposed a new generation of spectral
templates encompassing weak, intermediate, and strong first-order phase
transitions, with the strong regime ($\alpha\sim0.5$) constructed for the first
time. While the weak-transition limit is well described by earlier results,
intermediate and strong transitions exhibit deviations from the conventional
linear scaling with the source duration. For strongly supercooled transitions
proceeding in a vacuum-energy--dominated background, we adopt the framework of
Ref.~\cite{Caprini:2024hue},
\begin{equation}
h^2\Omega_{\rm GW}(f)=
h^2\Omega_p
\left(\frac{f}{f_p}\right)^{n_1}
\left[1+\left(\frac{f}{f_p}\right)^{n_1-n_2}\right]^{-1},
\label{eq:strong_pt_temp}
\end{equation}
with slopes
\bea
n_1=2.4 \quad (f<f_{\rm bub}),
\qquad
n_2=-2.4 \quad (f>f_{\rm bub}),
\eea
where $f_{\rm bub}$ is the peak frequency associated with bubble collisions. The
mapping between thermodynamic and spectral quantities is
\bea
h^2\Omega_p = h^2F_{\rm GW,0}\,
A_{\rm str}\,\tilde K^2
\left(\frac{H_*}{\beta}\right)^2,
\eea
\bea
\label{eq:fbub}
f_{\rm bub} \simeq 0.11\,h_{*,0}
\left(\frac{\beta}{H_*}\right),
\eea
with
\bea
\tilde K=\frac{\alpha}{1+\alpha},
\qquad
A_{\rm str}\simeq0.05,
\eea
and
\bea
h^2F_{\rm GW,0}\simeq1.64\times10^{-5}
\left(\frac{100}{g_*}\right)^{1/3}.
\eea

\medskip
Finally, the bubble-wall velocity $v_w$ is determined either macroscopically from
hydrodynamic matching or dynamically once plasma friction is included
\cite{Steinhardt:1981ct,Moore:1995ua,John:2000zq,Espinosa:2010hh,Bodeker:2009qy}.
Assuming local thermal equilibrium and an infinitesimally thin wall, the plasma is
described by a perfect-fluid energy--momentum tensor
$T^{\mu\nu}=(\rho+p)u^\mu u^\nu+pg^{\mu\nu}$.
Energy--momentum conservation across the wall yields the Rankine--Hugoniot
conditions
\begin{align}
w_+ \gamma_+^2 v_+ &= w_- \gamma_-^2 v_- , \\
w_+ \gamma_+^2 v_+^2 + p_+ &= w_- \gamma_-^2 v_-^2 + p_- ,
\end{align}
with $w=\rho+p$. These equations admit deflagration, detonation, and hybrid
solutions, but do not uniquely determine $v_w$, allowing instead a continuous
family of solutions.

Including deviations from equilibrium leads to a friction force acting on the
wall. The scalar background $\phi(z)$ satisfies
\begin{equation}
\phi''-\frac{\partial V_{\rm eff}}{\partial\phi}
+\sum_i\frac{\partial m_i^2}{\partial\phi}
\int\frac{d^3p}{(2\pi)^3 2E}\,\delta f_i=0,
\end{equation}
where $\delta f_i$ obey Boltzmann equations in the wall frame. In the
relaxation-time approximation, the resulting friction force is linear in the wall
velocity,
\begin{equation}
F_{\rm fric}=\eta\,v_w,
\qquad
\eta\simeq\sum_i\frac{g_i}{T}
\left(\frac{\partial m_i^2}{\partial\phi}\right)^2
\int dz\,\frac{\phi'(z)^2}{\Gamma_i},
\end{equation}
with $\Gamma_i$ the interaction rates. The terminal velocity follows from the
force-balance condition $\Delta V_{\rm eff}(T)=\eta v_w$. If friction cannot
balance the vacuum pressure, the wall accelerates into a runaway regime.

In practice, the bubble-wall velocity is obtained by either solving the
hydrodynamic matching equations with a suitable selection criterion or by
numerically solving the coupled scalar-field and Boltzmann equations. In this
work, we employ either approach depending on the validity conditions of
\texttt{WallGo}~\cite{Ekstedt:2024fyq,vandeVis:2025plm}, which we use for the
numerical determination of $v_w$.

\begingroup
\raggedright
\bibliographystyle{jhep}
\bibliography{biblio}
\endgroup
\end{document}